\documentclass [aps,pre,twocolumn,10pt,superscriptaddress,longbibliography] {revtex4-1}
\usepackage{graphicx}
\usepackage{subfigure}
\usepackage{amsmath}
\usepackage{amsfonts}
\usepackage{amssymb}
\usepackage{amsbsy}
\usepackage{xcolor}
\usepackage{siunitx}
\usepackage[colorlinks=true,allcolors=blue]{hyperref}
\usepackage{tikz}
\usepackage{scalerel}
\usepackage{xifthen}
\usepackage{chngcntr}
\usetikzlibrary{svg.path}

\AtBeginDocument{
    \newwrite\bibnotes
    \def\bibnotesext{Notes.bib}
    \immediate\openout\bibnotes=\jobname\bibnotesext
    \immediate\write\bibnotes{@CONTROL{%
    apsrev41Control,author="08",editor="1",pages="1",title="0",year="1"}}
     \if@filesw
     \immediate\write\@auxout{\string\citation{apsrev41Control}}%
    \fi
}

\definecolor{orcidlogocol}{HTML}{A6CE39}
\tikzset{
  orcidlogo/.pic={
    \fill[orcidlogocol] svg{M256,128c0,70.7-57.3,128-128,128C57.3,256,0,198.7,0,128C0,57.3,57.3,0,128,0C198.7,0,256,57.3,256,128z};
    \fill[white] svg{M86.3,186.2H70.9V79.1h15.4v48.4V186.2z}
                 svg{M108.9,79.1h41.6c39.6,0,57,28.3,57,53.6c0,27.5-21.5,53.6-56.8,53.6h-41.8V79.1z M124.3,172.4h24.5c34.9,0,42.9-26.5,42.9-39.7c0-21.5-13.7-39.7-43.7-39.7h-23.7V172.4z}
                 svg{M88.7,56.8c0,5.5-4.5,10.1-10.1,10.1c-5.6,0-10.1-4.6-10.1-10.1c0-5.6,4.5-10.1,10.1-10.1C84.2,46.7,88.7,51.3,88.7,56.8z};
  }
}

\newcommand\orcid[1]{\href{https://orcid.org/#1}{\mbox{\scalerel*{
\begin{tikzpicture}[yscale=-1,transform shape]
\pic{orcidlogo};
\end{tikzpicture}
}{|}}}}

\begin{document}

\title{Oscillatory rheotaxis of active droplets in microchannels}

\author{Ranabir Dey\orcid{0000-0002-0514-7357}}
\email{ranabir@mae.iith.ac.in}
\affiliation{Dynamics of Complex Fluids, Max Planck Institute for Dynamics and Self-Organization, Am Fa\ss{}berg 17, 37077 G\"ottingen, Germany
and Institute for the Dynamics of Complex Systems, Georg August Universit\"at G\"ottingen, Germany}
\affiliation{Department of Mechanical and Aerospace Engineering, Indian Institute of Technology Hyderabad, Kandi, Sangareddy, Telengana- 502285, India}

\author{Carola M. Buness\orcid{0000-0003-0437-0538}} 
\affiliation{Dynamics of Complex Fluids, Max Planck Institute for Dynamics and Self-Organization, Am Fa\ss{}berg 17, 37077 G\"ottingen, Germany
and Institute for the Dynamics of Complex Systems, Georg August Universit\"at G\"ottingen, Germany}

\author{Babak Vajdi Hokmabad\orcid{0000-0001-5075-6357}}
\affiliation{Dynamics of Complex Fluids, Max Planck Institute for Dynamics and Self-Organization, Am Fa\ss{}berg 17, 37077 G\"ottingen, Germany
and Institute for the Dynamics of Complex Systems, Georg August Universit\"at G\"ottingen, Germany}

\author{Chenyu Jin\orcid{0000-0002-5552-0340}}
\affiliation{Dynamics of Complex Fluids, Max Planck Institute for Dynamics and Self-Organization, Am Fa\ss{}berg 17, 37077 G\"ottingen, Germany
and Institute for the Dynamics of Complex Systems, Georg August Universit\"at G\"ottingen, Germany}
\affiliation{Physics Department, University of Bayreuth, 95440 Bayreuth, Germany}

\author{Corinna C. Maass\orcid{0000-0001-6287-4107}}
\email{c.c.maass@utwente.nl}
\affiliation{Dynamics of Complex Fluids, Max Planck Institute for Dynamics and Self-Organization, Am Fa\ss{}berg 17, 37077 G\"ottingen, Germany
and Institute for the Dynamics of Complex Systems, Georg August Universit\"at G\"ottingen, Germany}
\affiliation{Physics of Fluids Group, Max Planck Center for Complex Fluid Dynamics, MESA+ Institute and J. M. Burgers Center for Fluid Dynamics, University of Twente, PO Box 217,7500 AE Enschede, The Netherlands}

\begin{abstract}
Biological microswimmers are known to navigate upstream of an external flow (positive rheotaxis) in trajectories ranging from linear, spiral to oscillatory. Such rheotaxis stems from the interplay between the motion and complex shapes of the microswimmers, e.g. the chirality of the rotating flagella, the shear flow characteristics, and the hydrodynamic interaction with a confining surface. Here, we show that an isotropic, active droplet microswimmer exhibits a unique oscillatory rheotaxis in a microchannel despite its simple spherical geometry. The swimming velocity, orientation, and the chemical wake of the active droplet undergo periodic variations between the confining walls during the oscillatory navigation. Using a hydrodynamic model and concepts of dynamical systems, we demonstrate that the oscillatory rheotaxis of the active droplet emerges primarily from the interplay between the hydrodynamic interaction of the finite-sized microswimmer with all the microchannel walls, and the shear flow characteristics. Such oscillatory rheotactic behavior is different from the directed motion near a planar wall observed previously for artificial microswimmers in shear flows. Our results provide a realistic understanding of the behaviour of active particles in confined microflows, as will be encountered in majority of the applications like targeted drug delivery.   
\end{abstract}
\maketitle
\section*{Introduction}
Microorganisms navigate upstream of external flows in confined environments ranging from biological, like sperm cells in the reproductive tract \cite{kantsler2014rheotaxis} and bacteria (e.g. \textit{E. coli}) in the upper urinary tract \cite{lane2005role}, to medical, like  \textit{E. coli} in catheters \cite{figueroa2020coli}. 
Classically, the locomotion of organisms in shear flows by continuous reorientation in response to changes in external velocity gradients is called rheotaxis \cite{bretherton1961rheotaxis}. 
Microorganisms, like bacteria, exhibit positive rheotaxis, i.e. swim upstream, due to a combination of passive mechanisms stemming from the interplay between the complex shape and motion of the microorganisms, activity-induced hydrodynamic interaction with the confining surfaces in the vicinity, and the characteristics of the ambient flow \cite{kaya2012direct,uppaluri2012_flow,fu2012bacterial,figueroa2015living,junot2019swimming,mathijssen2019oscillatory,jing2020chirality}. 
Specifically for bacteria, the interaction of the intricately shaped (chirality), moving flagella with the shear flow plays a critical role in dictating the rheotactic trajectory \cite{fu2012bacterial,mathijssen2019oscillatory,jing2020chirality}. 
Note then that in the absence of asymmetrical and/or counter-rotating components, like flagella, most of the mechanisms for passive rheotaxis are inapplicable.
Therefore, the rheotactic characteristics of artificial microswimmers, e.g. Janus particles and active droplets, cannot be predicted \textit{a priori} solely based on the understanding of rheotaxis of biological microswimmers.

In recent years, the scientific community has begun to address the immense potential of artificial microswimmers in applications like targeted cargo delivery and water remediation \cite{soler2014catalytic,gao2014environmental,ren2017rheotaxis}.
In the majority of these applications, the artificial microswimmers are inevitably required to navigate external flows in confinements, i.e. to rheotax. 
Despite this fact, quantitative understanding of rheotaxis of artificial microswimmers is surprisingly limited. 
Only recently it was demonstrated that spherical Janus particles near a surface exhibit robust cross-stream migration at a definite orientation relative to a uniform shear flow \cite{katuri2018cross,uspal2015rheotaxis}. 
Furthermore, the rheotactic characteristics of gold-platinum Janus rods was shown to depend on the interfacial gold/platinum length ratio \cite{brosseau2019relating} and the shear flow strength \cite{baker2019fight}. 
An analysis of these few works reveals three things- one, the prevailing understanding of rheotaxis of artificial microswimmers is limited to their observed behaviour near a surface and not in a confinement where simultaneous interaction with multiple surfaces in presence of a non-uniform shear flow is possible; two, a physical understanding of the characteristics of the associated chemical trail for artificial microswimmers during rheotaxis is still lacking; and three, the present understanding is solely based on the behaviour of Janus particles, and exclude active droplets which are an important class of artificial microswimmers. Active droplets are essentially oil/aqueous droplets in aqueous/oil surfactant medium which are intrinsically symmetric, unlike Janus particles \cite{maass2016_swimming,izri2014self}. 
Active droplets spontaneously break the symmetry of the isotropic base state to self-propel using Marangoni stresses \cite{morozov2019nonlinear, hokmabad2021emergence}. 
Hence, these are not only simpler to generate and manipulate compared to Janus particles, but are also biocompatible. 
All these features make active droplets the foremost candidate for majority of the envisioned applications for artificial microswimmers, specifically targeted drug delivery. 
Therefore, rheotaxis of active droplets in confinements in itself deserves a detailed study. 
Here, narrow channels are of particular interest, since this geometry applies to almost all conceivable problems in nature or technological application. 

Here, we investigate the rheotaxis of active droplets in a non-uniform, Poiseuille type shear flow in a quasi-2D microchannel. 
We demonstrate that the active droplet swims upstream in a surprising, steady oscillatory trajectory despite its inherent spherical symmetry. 
As it oscillates upstream, the active droplet accelerates and decelerates, and continuously changes its swimming orientation in a periodic manner between the two confining walls perpendicular to the plane of oscillation. 
We also implement a fluorescent microscopy technique to reveal the physical characteristic of the chemical (filled micelle) trail of the active droplet during such oscillatory rheotaxis. 
Using a semi-analytical hydrodynamic model, we show that the hydrodynamic interaction of the finite-sized, pusher-type microswimmer with all the walls of the microchannel, along with the shear flow characteristics, dictate the essential features of the oscillatory rheotactic dynamics. 
The oscillatory rheotaxis of active droplets demonstrated here reveals a new paradigm for rheotaxis of artificial microswimmers, beyond the linear motion at a definite orientation to a uniform shear flow demonstrated so far \cite{katuri2018cross,baker2019fight,ren2017rheotaxis}. 
Furthermore, we believe that our understanding of this new rheotactic behaviour will aid in conceptualizing and designing practical applications like targeted drug delivery which involve navigation of microswimmers in confinements.       
\section*{Results}
\subsection*{Oscillatory rheotaxis of an active droplet}
\begin{figure*}
\centering
\includegraphics[width=\textwidth]{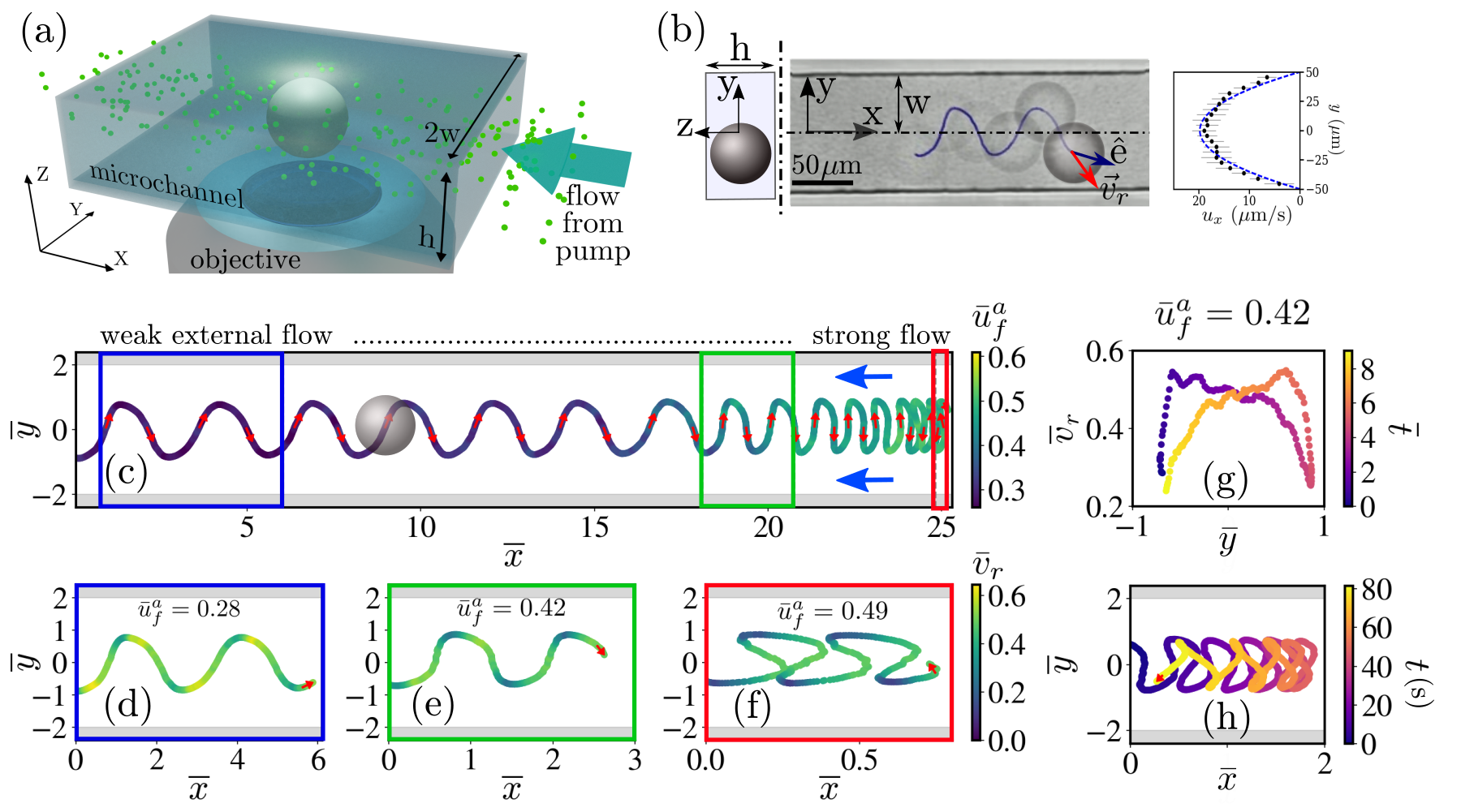}
\caption{\label{fig1:rheotaxis_traj} \small{\textbf{Oscillatory upstream (positive) rheotaxis of an active droplet in a microchannel for increasing strength of an imposed pressure-driven flow.} \textbf{(a)} Microscopy setup schematic showing the active droplet in a quasi-2D microchannel with height $\approx 52$ $\mu$m, which is similar to the droplet diameter ($2R_d \approx 50$ $\mu$m). \textbf{(b)} A time-lapse image, constructed using bright field microscopy images, showing the active droplet swimming upstream in an oscillatory path. The droplet's centroid trajectory is shown with the solid blue line. The experimentally obtained (using PIV) velocity profile in the plane of oscillation for the imposed pressure-driven flow  is also shown (markers: experimental; dotted line: quadratic fit). \textbf{(c)} Variation in the oscillatory rheotactic trajectory of the active droplet with increasing average velocity of the imposed flow. The trajectory is colour coded by the non-dimensional average flow velocity ($\bar{u}_f^a=u_f^a/v_0$; $v_0$ is the intrinsic droplet swimming velocity) in the plane of oscillation $(\bar{x}=x/R_d; \bar{y}=y/R_d)$. \textbf{(d)-(f)} Variation in the non-dimensional translational or rheotactic velocity $(\bar{v}_r=v_r/v_0)$  over the oscillatory trajectory for increasing values of $\bar{u}_f^a$. \textbf{(g)} Variation in  $\bar{v}_r$ with the transverse location of the active droplet over one wavelength of the oscillation for $\bar{u}_f^a=0.42$. The variation in $\bar{v}_r$ is colour coded by the non-dimensional time ($\bar{t}$). \textbf{(h)} Trapping of the oscillating active droplet, at sufficiently high $\bar{u}_f^a$, within a region comparable to its diameter for over a minute.}}  
\end{figure*}
In this work, oil droplets (CB 15; dia. $2R_d \approx 50$ $\mu$m) slowly dissolving in supramicellar aqueous solution of ionic surfactant (7.5 wt.$\%$ aqueous solution of TTAB) comprise the active droplets. 
We observe the swimming behaviour of such self-propelling droplets in a quasi-2D microchannel (width $2w \approx \SI{100}{\um}$; height $h \approx 2R_d \approx \SI{52}{\um}$) in presence of an external pressure-driven flow using bright field microscopy (Fig. \ref{fig1:rheotaxis_traj}a). 
The microchannel is fabricated from PDMS using photolithography and softlithography techniques, while the pressure-driven flow is actuated and maintained using a syringe pump. 
We study the swimming dynamics of the active droplet in the microchannel for different imposed flow rates ($\sim$\SIrange{40}{94} {\pico\litre\per\second}), represented here by the corresponding average flow velocity ($u_f^a \approx$\SIrange{7.7}{18.1}{\um\per\second}) in the X-Y plane (Fig. \ref{fig1:rheotaxis_traj}a). 
$u_f^a$ is evaluated using PIV analysis, for which the aqueous surfactant solution is seeded with tracer particles.
The order of magnitude of the resultant shear is $O(\dot{\gamma})\sim u_f^a/w \sim 0.15-0.36$ s$^{-1}$. 
The variations in the trajectory, rheotactic or translational velocity (i.e. the magnitude of the tracked velocity vector $|\vec{v}_r| \equiv v_r$), and the intrinsic orientation of the active droplet (defined by the orientation of the intrinsic swimming velocity vector $\vec{v}_0=v_0 \hat{e}$) in the external flow are extracted from the bright-field microscopy images using in house Python code (see the Methods section for the experimental and the post-processing details). 
In the subsequent discussions, spatial variables are non-dimensionalized by $R_d$ (particularly, $\bar{h}=2$), velocities by the intrinsic swimming speed   of the active droplet in an unbounded quiescent medium, $|\vec{v}_0| \equiv v_0 \approx 29.5$ $\mu$m/s, and times by $R_d/v_0$ . 
All such non-dimensionalized quantities are denoted by overbars.

\begin{figure}
\centering
\includegraphics[width=0.5\textwidth]{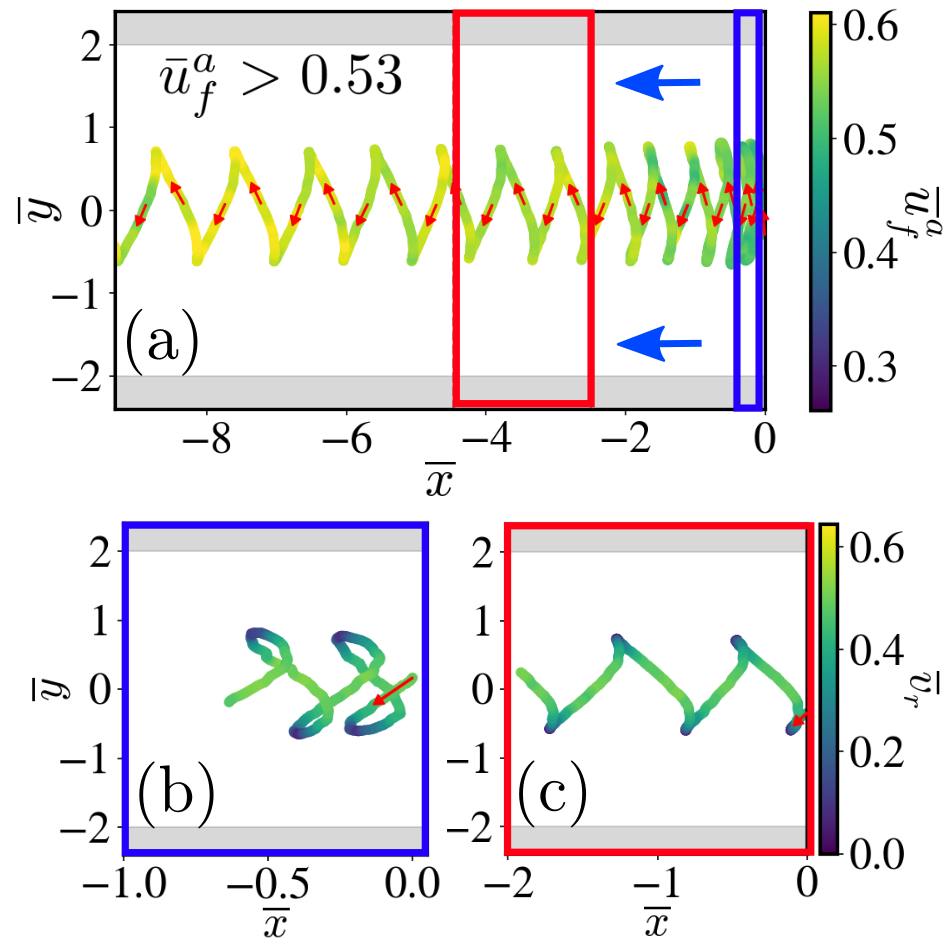}
\caption{\label{fig2:drifting_traj} \textbf{Downstream drift of the active droplet in a swinging trajectory on increasing $\bar{u}_f^a$ beyond a threshold.} \textbf{(a)} Variation in the swinging trajectory of the active droplet, as it drifts downstream, with increasing $\bar{u}_f^a$ beyond a threshold value ($\sim 0.53$). The trajectory is colour coded by $\bar{u}_f^a$. \textbf{(b)-(c)} Variation in $\bar{v}_r$ over the swinging trajectory for increasing values of $\bar{u}_f^a$.}
\end{figure}

It is worth noting first that in absence of the external flow, an active droplet self-propels in the quasi-2D microchannel following a linear trajectory while adhering to one of the side walls. 
On actuating the pressure-driven flow, an identical active droplet in the microchannel swims upstream in a novel oscillatory trajectory in the X-Y plane (Fig. \ref{fig1:rheotaxis_traj}b). 
With increasing $\bar{u}_f^a$, the active droplet exhibits persistent upstream oscillation, but the wavelength of the oscillatory trajectory gradually reduces (Fig. \ref{fig1:rheotaxis_traj}c). 
At a definite value of $\bar{u}_f^a$, the non-dimensional rheotactic velocity $(\bar{v}_r)$ of the active droplet varies along the trajectory in a periodic manner (Fig. \ref{fig1:rheotaxis_traj}d-f).
Specifically, $\bar{v}_r$ reduces sharply as the droplet approaches a side wall (Fig. \ref{fig1:rheotaxis_traj}g). 
However, in the immediate vicinity of the side wall, the droplet accelerates and $\bar{v}_r$ reaches its maximum value.
Thereafter, $\bar{v}_r$ gradually reduces as the droplet swims towards the opposite side wall while advancing upstream. 
Eventually, the droplet again sharply decelerates after it crosses the channel center-line and approaches that wall.
At moderately high value of $\bar{u}_f^a$, the droplet eventually fails to effectively swim upstream.
Interestingly, it is possible to trap the oscillating droplet within a small region by judiciously tuning $\bar{u}_f^a$ about a threshold value ($\sim 0.53$) (Fig. \ref{fig1:rheotaxis_traj}h).
Finally, beyond $\bar{u}_f^a \sim 0.53$, the active droplet drifts downstream in a swinging trajectory (Fig. \ref{fig2:drifting_traj}).

There are three things to be noted about the downstream drift of the active droplet. 
First, the active droplet drifts downstream for $\bar{u}_f^a$ relatively smaller than 1, contrary to earlier theoretical works which predicted downstream drift of microswimmers for $\bar{u}_f^a > 1$ \cite{zottl2012_nonlinear, stark2016swimming}. 
Second, the wavelength of the swinging drift trajectory increases with increasing $\bar{u}_f^a$ (Fig. \ref{fig2:drifting_traj}a), which is opposite to the trend observed during the oscillatory upstream rheotaxis (Fig. \ref{fig1:rheotaxis_traj}c). 
Finally, third, the periodic variation in $\bar{v}_r$ over the swinging downstream trajectory remains qualitatively similar to that observed during upstream oscillation (Fig. \ref{fig2:drifting_traj}b-c).
\subsection*{Variations in the intrinsic and translational orientations of the active droplet}
\begin{figure*}
\centering
\includegraphics[width=\textwidth]{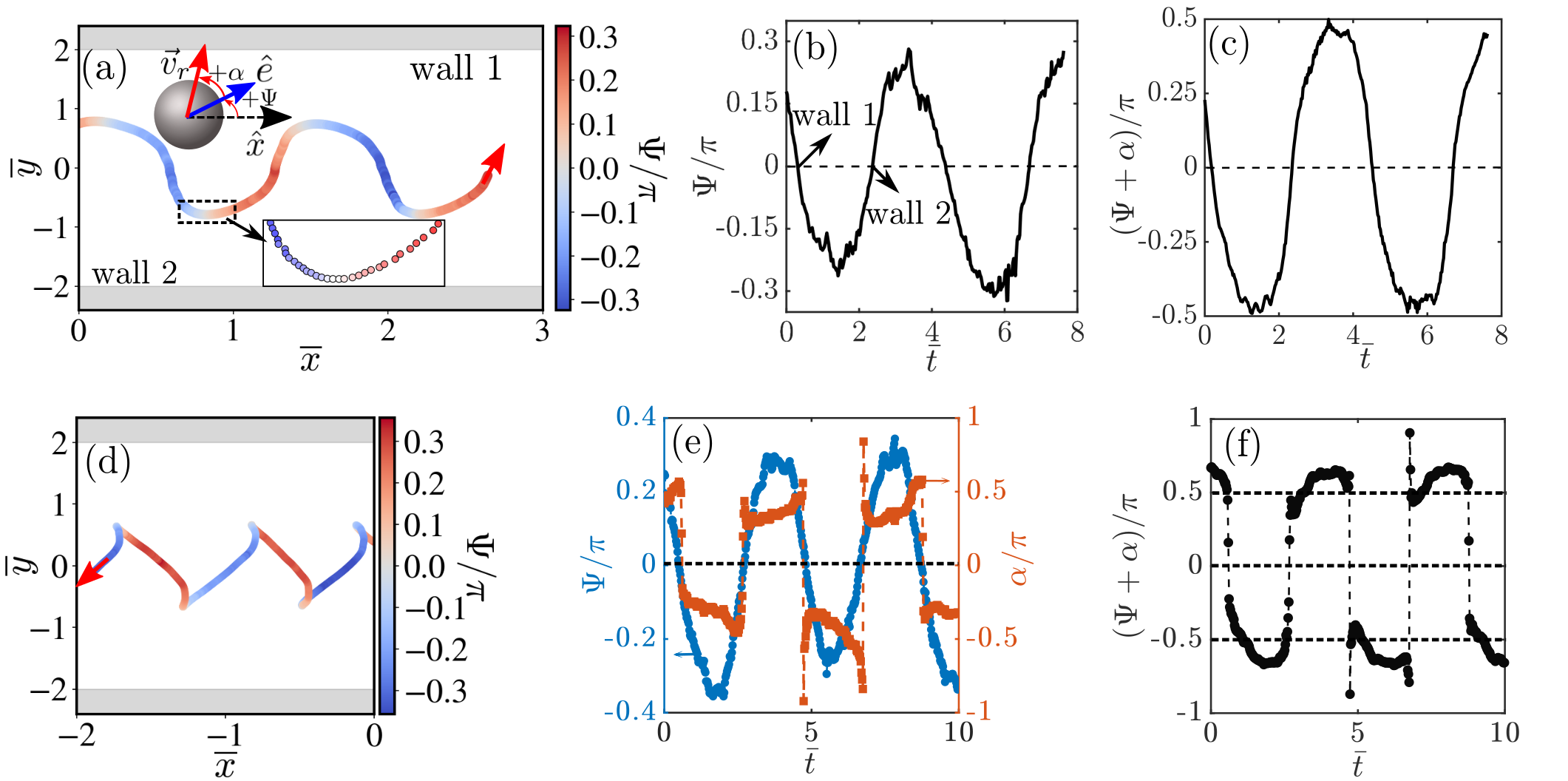}
\caption{\label{fig3:orientation} \textbf{Variations in swimming and translational orientations of the active droplet during upstream rheotaxis and downstream drift.} \textbf{(a)} Variation in the intrinsic orientation of the active droplet over the oscillatory rheotactic trajectory for $\bar{u}_f^a=0.42$. The swimming orientation is defined by the angle $\Psi$ between the orientation vector $(\hat{e})$ and the (positive) horizontal unit vector $(\hat{x})$ as shown in the schematic. Positive (negative) values of $\Psi$ represent anti-clockwise (clockwise) orientation about $+\hat{x}$. \textbf{(b)} Periodic variation in $\Psi$ with the non-dimensional time $(\bar{t})$ during upstream rheotaxis. \textbf{(c)} Variation in the orientation of the translational direction (direction of the rheotactic/tracked velocity vector  $\vec{v}_r$; see schematic in a) relative to $+\hat{x}$, as given by $(\alpha+\Psi)$, with $\bar{t}$.\textbf{(d)} Variation in $\Psi$ over the swinging trajectory during downstream drift of the active droplet. \textbf{(e)} Variations in $\Psi$ and $\alpha$, and \textbf{(f)} variation in $(\alpha+\Psi)$ with $\bar{t}$ during the downstream drift. Note that the active droplet is always oriented upstream $(|\Psi|/\pi<0.5)$ even as it drifts downstream.}
\end{figure*}
To further quantify the oscillatory rheotactic characteristics, we extract the variation in the intrinsic orientation $(\hat{e})$ of the active droplet.
This is shown here by the variation in the angle $\Psi$ which defines the orientation of the unit vector $\hat{e}$ relative to $+\hat{x}$ in the X-Y plane $(\hat{e}=\cos{\Psi} \hat{x}+ \sin{\Psi} \hat{y})$ (schematic in Fig. \ref{fig3:orientation}a).
Here, $\Psi \in \{-\pi,\pi \}$, and $\Psi$ is positive (negative) when measured in an anti-clockwise (clockwise) sense from $+\hat{x}$ (schematic in Fig. \ref{fig3:orientation}a).
$\Psi$ varies in a periodic manner over the oscillatory rheotactic trajectory (Fig. \ref{fig3:orientation}a).
As the active droplet approaches a side wall with sharply decreasing $\bar{v}_r$ (Fig. \ref{fig1:rheotaxis_traj}g), it continuously orients itself to become parallel to that wall ($|\Psi|$ reduces; Fig. \ref{fig3:orientation}a, b). 
Eventually, the droplet microswimmer becomes parallel to the wall ($\Psi=0$) in its immediate vicinity for a brief period (residence time $\sim \SI{0.17}{\second}$ at $|\Psi|<0.1\pi$; Fig. \ref{fig3:orientation}a inset, b).
Subsequently, as the droplet accelerates (Fig. \ref{fig1:rheotaxis_traj}g), it simultaneously orients away from the adjacent wall and towards the opposite side wall ($|\Psi|$ increases; Fig. \ref{fig3:orientation}a, b).
Even when the droplet decelerates (Fig. \ref{fig1:rheotaxis_traj}g), it continues to orient itself towards the opposite wall, and crosses the channel center-line with the maximum cross-stream orientation (Fig. \ref{fig3:orientation}b).
Thereafter, the droplet begins to approach the other wall with decreasing $|\Psi|$ as before (Fig. \ref{fig3:orientation}b).

Note that the direction of cross-stream migration of the active droplet is different from its intrinsic orientation. 
This is expressed here by additionally defining an offset angle $\alpha$ between $\hat{e}$ and $\vec{v}_r$ (schematic in Fig. \ref{fig3:orientation}a).
The variation in the translational direction relative to $+\hat{x}$ is then given by the angle $\Psi+\alpha$ (Fig. \ref{fig3:orientation}c).
$\Psi+\alpha$ varies smoothly about $0$ in the vicinity of the side walls (Fig. \ref{fig3:orientation}c). 
This implies that the translational orientation of the active droplet becomes truly parallel to the side walls in their immediate vicinity, and the droplet does not crash into the side walls during the oscillatory upstream rheotaxis.

We also evaluate the variations in the intrinsic and translational orientations of the active droplet during the swinging downstream drift  (Fig. \ref{fig3:orientation}d-f).
Interestingly, the droplet microswimmer always remains oriented upstream $(|\Psi|/\pi < 0.5)$ even as it moves downstream (Fig. \ref{fig3:orientation}d, e).
However, for $\bar{u}_f^a > 0.53$, the resulting velocity field makes $\vec{v}_r$ orient downstream $(|\Psi+\alpha|/\pi>0.5)$ as the droplet traverses the width of the microconfinement (Fig. \ref{fig3:orientation}f).
Note that while $\Psi$ varies smoothly even during the downstream motion, $\alpha$ undergoes sharp variations near the side walls (Fig. \ref{fig3:orientation}e). 
The latter is reflected in the variation of $\Psi+\alpha$ (Fig. \ref{fig3:orientation}f), and facilitates the downstream motion of the active droplet.   
In essence, it can be concluded that the active droplet does not self-propel or swim downstream, but it is pushed downstream in a swinging trajectory for $\bar{u}_f^a > 0.53$.
\subsection*{Orientation of the chemical field of the active droplet during oscillatory rheotaxis}
\begin{figure}
\centering
\includegraphics[width=1\columnwidth]{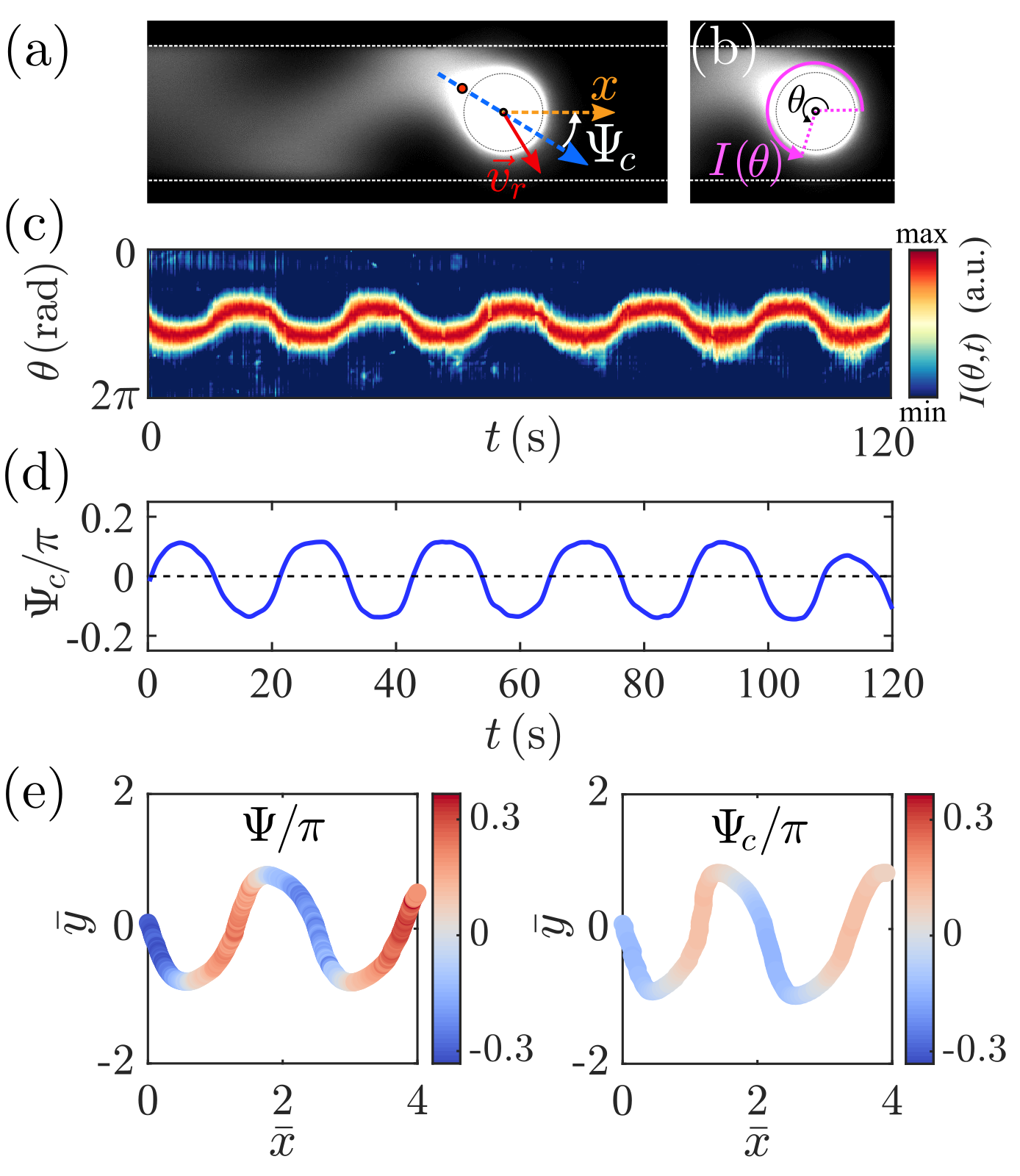}
\caption{\label{fig5:psi_c} \textbf{Characteristics of the chemical (filled micelle) trail of the active droplet during oscillatory rheotaxis.} \textbf{(a)} A fluorescence microscopy image showing the filled micelle trail of the active droplet during the oscillatory rheotaxis. The active droplet is doped with the fluorescent dye Nile Red. The dye co-migrates with the oil molecules into the filled micelles thereby making these fluoresce when illuminated. \textbf{(b)} defines the calculation of the fluorescence intensity $(I)$ in the immediate vicinity of the droplet interface. The location of the point of interest is defined by the angle $\theta$ relative to $+\hat{x}$. \textbf{(c)} Kymograph of $I(\theta)$. (d) Temporal variation of the orientation of the filled micelle tail of the active droplet, as defined by $\Psi_c$ which is supplementary to the orientation of the point of maximum $I$ relative to $+\hat{x}$ (see (a)). \textbf{(e)} Comparison between the variations in the intrinsic orientation of the active droplet $(\Psi)$ and $\Psi_c$ over the oscillatory rheotactic trajectory for $\bar{u}_f^a \sim 0.42$. The chemical tail is closely aligned with the swimming orientation, and is not significantly distorted by the resulting flow field.}
\end{figure}
Based on our previous work on the motility of active droplets, we know that the chemical (filled micelle) field generated by the active droplet can locally interact with it altering the resulting orientation \cite{hokmabad2021emergence}.
It is then only logical to wonder whether similar secondary chemical interactions play any role in the above described rheotactic dynamics of the active droplet.
To this end, we implement a fluorescence microscopy technique to visualize the filled micelle trail of the active droplet during the oscillatory upstream rheotaxis (Fig. \ref{fig5:psi_c}a; see Methods section for details). 
Additionally, this endeavour also provides insight about the behaviour of the chemical trail during rheotaxis of artificial microswimmers in general; an aspect which has not been probed before.

We extract the fluorescence intensity profile $I(\theta)$ emitted by the filled micelle field in the vicinity of the droplet interface over time (Fig. \ref{fig5:psi_c}b), and map it onto a kymograph (Fig. \ref{fig5:psi_c}c).
The kymograph (Fig. \ref{fig5:psi_c}c) shows that the orientation of the filled micelle trail of the droplet also follows a periodic variation during the oscillatory upstream rheotaxis.
To quantify this orientational variation of the filled micelle tail, we define an angle $\Psi_c$ which is supplementary to the extracted orientation angle of $[I(\theta)]_{max}$ relative to $+\hat{x}$ (i.e. $\Psi_c=\pi-\theta_{max}$; Fig. \ref{fig5:psi_c}a).
Fig. \ref{fig5:psi_c}d quantitatively shows the periodic variation of the orientation of the filled micelle trail during the oscillatory rheotaxis.
Here, we assume  that the location of $[I(\theta)]_{max}$ approximately coincides with the location of the maximum filled micelle concentration, which is at the posterior stagnation point (albeit in the droplet reference frame) \cite{hokmabad2021emergence}.
Therefore, under quiescent conditions, the swimming orientation $\Psi$ and the `chemical angle' $\Psi_c$  are constant and identical,  $\Psi=\Psi_c = \text{const.}$,  as the orientation of the active droplet coincides with the anterior stagnation point.
Therefore, for the present scenario, any offset between the variations in $\Psi$ and $\Psi_c$ will give an indirect measure of the distortion of the filled micelle tail orientation due to the resulting flow field during rheotaxis within the microconfinement; such distortions can trigger changes in the intrinsic swimming orientation \cite{hokmabad2021emergence}.
However, a comparison between $\Psi$ and $\Psi_c$ over the rheotactic trajectory (Fig. \ref{fig5:psi_c}e) shows that these are satisfactorily similar.
Therefore, it can be concluded that the filled micelle trail is not significantly distorted by the resulting velocity field. 
Accordingly, it is safe to assume that the filled micelle field does not additionally interfere with the rheotactic dynamics via local chemical interactions.
\section*{Understanding the oscillatory rheotaxis based on hydrodynamic interaction}
\begin{figure*}
\centering
\includegraphics[width=\textwidth]{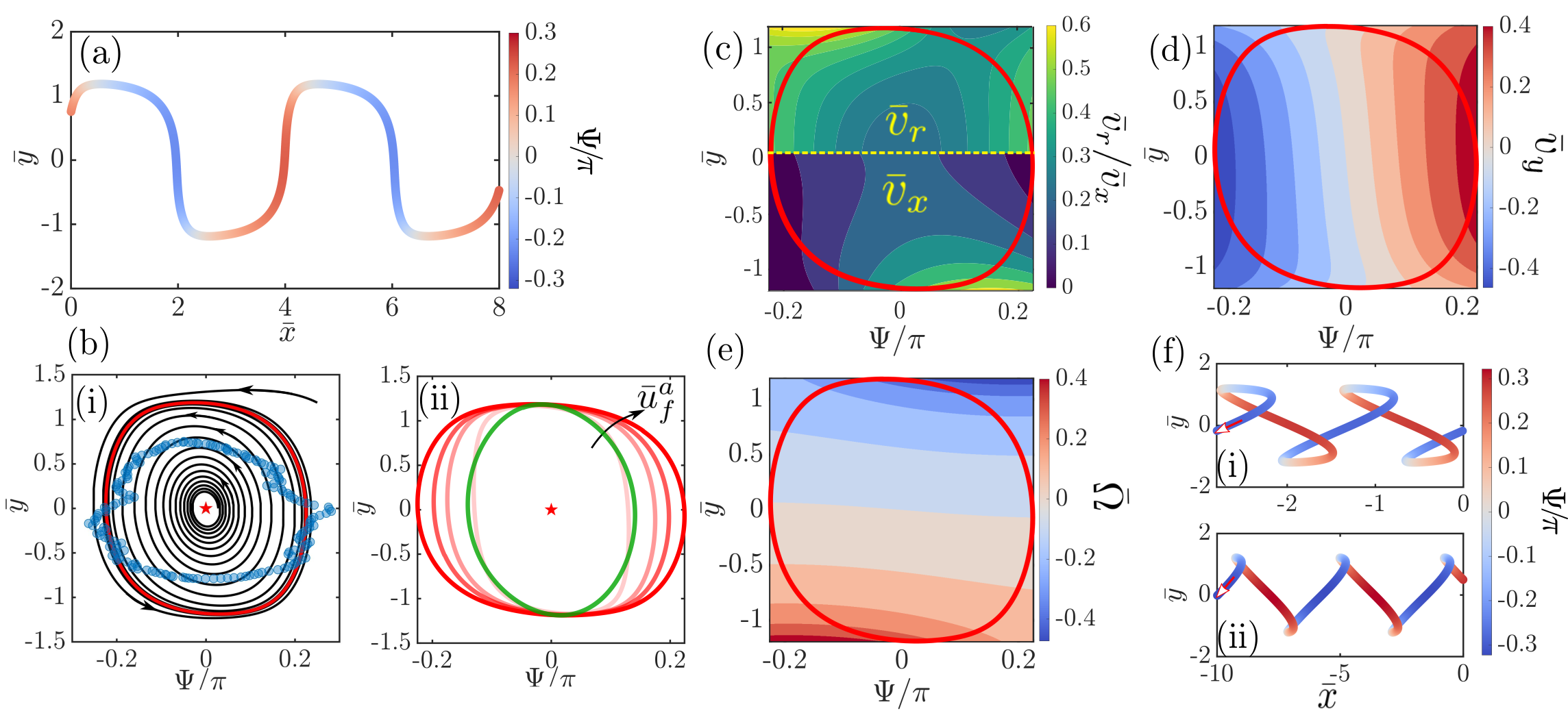}
\caption{\label{fig6:theory} \textbf{Dynamics of the oscillatory rheotaxis of the active droplet captured using a hydrodynamic model.} \textbf{(a)} Variation in the swimming orientation $(\Psi)$ of the active droplet over the oscillatory trajectory for $\bar{u}_f^a=0.42$ evaluated using Eqs. (\ref{eq_vxdot})-(\ref{eq_psidot}). \textbf{(b)(i)} Trajectories in the $\Psi$-$\bar{y}$ phase space for the upstream rheotaxis of the active droplet at $\bar{u}_f^a=0.42$. The microswimmer is initialized with a non-zero upstream orientation from the channel center as well as from near the microchannel wall. In both the cases the microswimmer approaches the identical closed trajectory (a stable limit cycle; solid red line). The blue markers represent the experimentally obtained stable trajectory. \textbf{(b)(ii)} Variation of the stable limit cycle with $\bar{u}_f^a$. Increasing opacity of the red line represents increasing $\bar{u}_f^a$, with the solid red line representing $\bar{u}_f^a=0.42$. The solid green line represents the stable limit cycle when the finite size effects of the microswimmer are neglected ($\beta=0$, $\bar{u}_f^a=0.42$). \textbf{(c)} The stable trajectory of the droplet microswimmer in the $\Psi$-$\bar{y}$ phase space  for $\bar{u}_f^a=0.42$ against the contour plots representing $\bar{v}_r$ (top half) and the non-dimensional x-component of the rheotactic velocity $(\bar{v}_x)$ (bottom half). \textbf{(d)} and \textbf{(e)} The stable trajectory of the droplet microswimmer against the countour plot representing the non-dimensional y-component of the rheotactic velocity $(\bar{v}_y)$ and the non-dimensional angular velocity of the microswimmer $(\bar{\Omega})$ respectively. \textbf{(f)(i-ii)} Trajectories and variations in orientation during the downstream drift for increasing values of $\bar{u}_f^a$ $(0.53<\bar{u}_f^a<1)$ evaluated using Eqs. (\ref{eq_vxdot})-(\ref{eq_psidot})}
\end{figure*}
 To understand the essential features of the novel oscillatory rheotaxis of the active droplet we develop a simplified, approximate hydrodynamic model.
 This endeavour stems from the understanding that for an axisymmetric, spherical microswimmer devoid of any chiral and/or counter-rotating components, like the active droplet, the mechanism for passive rheotaxis will likely stem from the interplay of the long-range hydrodynamic interaction with the confining walls \cite{spagnolie2012_hydrodynamics} and the change in swimming orientation due to the external shear flow (the Bretherton-Jeffery effect) \cite{bretherton1962motion}.
 In our simplified model, we neglect the coupling between the chemical concentration distribution and the hydrodynamics for the active droplet\cite{de2019flow}. 
 This to a certain extent is also justified by the conclusion that the filled micelle field does not interfere with the oscillatory dynamics.
 However, it is not obvious \textit{a priori} to what extent a purely hydrodynamic model will be able to capture the oscillatory rheotactic dynamics.
 
 We describe the droplet microswimmer by an equivalent squirmer model comprising of a superposition of a force dipole (\textit{`fd'}), a source dipole (\textit{`sd'}), and a source quadrupole (\textit{`sq'}) \cite{de2019flow, kuron2019hydrodynamic}.
 Accordingly, the velocity field of the swimming droplet can be written as \[\begin{split}\vec{u}_s&=
 \alpha \vec{u}_{fd}\left(\vec{r}-\vec{r}_0;\hat{e}(\Psi)\right)\\
 &+\beta \vec{u}_{sd}\left(\vec{r}-\vec{r}_0;\hat{e}(\Psi)\right)+\gamma \vec{u}_{sq}\left(\vec{r}-\vec{r}_0;\hat{e}(\Psi)\right).\end{split}\]
 Here, $\alpha$, $\beta$, and $\gamma$ are the strengths of the force dipole, source dipole, and the source quadrupole singularities respectively, $\vec{r}$ is the position vector, and $\vec{r}_0$ is the location of the squirmer.
 Subsequently, we use the method of images \cite{spagnolie2012_hydrodynamics,kuron2019hydrodynamic,de2019flow,de2016understanding} to evaluate the velocity field $(\vec{u}_{HI})$ resulting from the hydrodynamic interaction of the squirmer with all the four walls of the quasi-2D microchannel i.e. the two side walls normal to the X-Y plane, and the top and bottom walls parallel to the X-Y plane. 
 The total external velocity field $\vec{u}_t$ can be then written as

 \begin{equation} \label{eq_flowfield}
 \begin{split}
 \vec{u}_t&=\vec{u}_s+\vec{u}_{HI}-\vec{u}_f\\
 \text{with}&\\
 \vec{u}_{HI} &= \alpha \sum \vec{u}_{fd}^*\left(\vec{r}-\vec{r}_0^*;\hat{e}(\Psi)\right)+\beta \sum \vec{u}_{sd}^* \left(\vec{r}-\vec{r}_0^*;\hat{e}(\Psi)\right)\\
 &+\gamma \sum \vec{u}_{sq}^* \left(\vec{r}-\vec{r}_0^*;\hat{e}(\Psi)\right)\\
 \vec{u}_f&=c_f \bar{u}_f^a \left[ 1-\left( \frac{\bar{y}}{\bar{w}} \right)^2 \right]\hat{x}\\
 \end{split}  
 \end{equation}
 Here, $\vec{u}_{fd/sd/sq}^*$ represents the velocity field due to the image system at a wall corresponding to a particular singularity, $\vec{r}_0^*$ is the location of the image singularities for a particular wall, $\sum$ represents the vector sum of $\vec{u}_{fd/sd/sq}^*$ for all the walls, $\vec{u}_f$ is the imposed pressure-driven velocity field, and $\bar{w}=w/R_d$.
 $\vec{u}_f$ is approximated here as a plane Poiseuille flow in the X-Y plane in accordance with the experimental observation (Fig. \ref{fig1:rheotaxis_traj}b).
 Furthermore, there are two additional approximations involved here. 
 First, Eq. (\ref{eq_flowfield}) is a far-field approximation, and hence ignores that the droplet radius and wall distance are of similar order. 
Second, $\vec{u}_t$ is approximate to the leading order as we consider a total of twelve image systems - images for force dipole, source dipole, and source quadrupole singularities at each of the four walls, instead of the infinite series of images that should be considered to satisfy the no-slip boundary condition at the walls \cite{mathijssen2016hydrodynamics}.
 
 We use Faxen's first law  for a force-free sphere \cite{leal2007_advanced, kuron2019hydrodynamic, spagnolie2012_hydrodynamics} to evaluate the translational velocity of the squirmer in response to $\vec{u}_t$. 
 Accordingly, the x- and y- components of the translational (rheotactic) velocity can be evaluated as 
\begin{equation} \label{eq_vxdot}
\begin{split}
\bar{v}_x&=\cos{\Psi}-1.4 \bar{u}_f^a \left[ 1- \left(\frac{\bar{y}_0}{\bar{w}}\right)^2 - \frac{1}{3\bar{w}^2} \right]\\
&+\alpha \left[\pm \frac{0.875}{(\bar{y}_0 \pm \bar{w})^2} \pm \frac{0.078125}{(\bar{y}_0 \pm \bar{w})^4}\right] \sin{2\Psi}\\
&+\beta \left[\mp \frac{1}{4(\bar{y}_0 \pm \bar{w})^3} \pm \frac{3}{16(\bar{y}_0 \pm \bar{w})^5}-\frac{17}{16}\right] \cos{\Psi}\\
&+\gamma \left[\mp \frac{3}{16(\bar{y}_0 \pm \bar{w})^4} \mp \frac{5}{128(\bar{y}_0 \pm \bar{w})^6}\right] \cos{\Psi}\\
\end{split}
\end{equation}
\begin{equation} \label{eq_vydot}
\begin{split}
\bar{v}_y&=\sin{\Psi}+\alpha \left[\mp \frac{3(1-3\sin^2{\Psi})}{8(\bar{y}_0 \pm \bar{w})^2} \pm \frac{(7-11\sin^2{\Psi})}{64(\bar{y}_0 \pm \bar{w})^4}\right]\\
&+\beta \left[\mp \frac{1}{(\bar{y}_0 \pm \bar{w})^3} \pm \frac{1}{8(\bar{y}_0 \pm \bar{w})^5}-\frac{17}{16}\right] \sin{\Psi}\\
&+\gamma \left[\pm \frac{3(3+7\sin^2{\Psi})}{16(\bar{y}_0 \pm \bar{w})^4} \mp \frac{5(11-7\sin^2{\Psi})}{128(\bar{y}_0 \pm \bar{w})^6}\right]\\
\end{split}
\end{equation}
Here, each of the terms with $\pm/\mp$ represent two terms due to the image system at the two side walls; the first and second signs represent the contribution from the image systems at wall 2 and wall 1 respectively (Fig. \ref{fig3:orientation}a).  
The constant $17/16$ represents the total and only contribution from the image systems  at the top and bottom walls. 
Finally, the temporal change in the intrinsic swimming orientation of the squirmer is given by $\dot{\hat{e}}=\vec{\bar{\Omega}} \times \hat{e}$ where $\vec{\bar{\Omega}}=\frac{1}{2} \nabla \times \left(\vec{u}_{HI}+\vec{u}_f \right)$ is the angular velocity of the squirmer according to Faxen's second law for a torque-free sphere \cite{spagnolie2012_hydrodynamics,kuron2019hydrodynamic}. 
Following this, the rate of change of the swimming orientation can be evaluated as
\begin{equation} \label{eq_psidot}
\begin{split}
\dot{\Psi}&=-c_f \bar{u}_f^a \left(\frac{\bar{y}_0}{\bar{w}}\right)^2 \mp \alpha \frac{3}{16(\bar{y}_0 \pm \bar{w})^3} \sin{2\Psi}\\
&\pm \beta \frac{3}{8(\bar{y}_0 \pm \bar{w})^4} \cos{\Psi} \pm \gamma \frac{3}{8(\bar{y}_0 \pm \bar{w})^5} \sin{2\Psi} \\
\end{split}
\end{equation}
Eqs. (\ref{eq_vydot}) and (\ref{eq_psidot}) form a coupled system of equations, which is solved numerically for $\bar{y}_0(\bar{t})$ and $\Psi(\bar{t})$. 
Thereafter, Eq. (\ref{eq_vxdot}) is numerically integrated to evaluate the corresponding $\bar{x}_0(\bar{t})$. 
Fig. \ref{fig6:theory}a shows the trajectory of the squirmer colour-coded by $\Psi$ for $\bar{u}_f^a=0.42$, as obtained from Eqs. (\ref{eq_vxdot})-(\ref{eq_psidot}).
A comparison with Fig. \ref{fig3:orientation}a shows that the hydrodynamic model indeed satisfactorily captures the oscillatory rheotaxis of the active droplet (for $\alpha=0.20$, $\beta=0.22$, and $\gamma=-0.08$).
For a better understanding, we plot the rheotactic dynamics in the $\Psi$-$\bar{y}$ phase space.
\subsection*{Phase portraits}
The steady upstream oscillation of the active droplet represents a stable limit cycle in the $\Psi$-$\bar{y}$ phase space (Fig. \ref{fig6:theory}b(i); solid red line: theory; transparent blue markers: experiment).
Note that perfect upstream orientation at the microconfinement centerline ($\Psi=0,\bar{y}=0$; star marker in Fig. \ref{fig6:theory}b(i)) represents an unstable fixed point.
With increasing $\bar{u}_f^a$, the active droplet undergoes the self-sustaining upstream oscillation with sharper changes in $\Psi$ over $\bar{y}$ (Fig. \ref{fig6:theory}b(ii)).
This stems from the increase in droplet angular velocity $\bar{\Omega}$ with increasing angular velocity (vorticity) of the background flow, due to the one-to-one relationship between the two quantities following Faxen's second law.  
Additionally, $\bar{v}_x$ reduces with increasing $\bar{u}_f^a$ (see Eq. (\ref{eq_vxdot})).
The increase in $\bar{\Omega}$ and the simultaneous decrease in $\bar{v}_x$ culminate in the reducing oscillation wavelength with increasing $\bar{u}_f^a$ (Fig. \ref{fig1:rheotaxis_traj}c).
Before proceeding further, it is important to note here that the finite size of the active droplet relative to the microchannel, as embodied by $\beta$, plays a pivotal role in the observed rheotactic dynamics.
Due to its finite radius, the droplet consistently turns away from the side walls of the microchannel faster (higher $\bar{\Omega}$) resulting in enhanced variation in $\Psi$ (compare the opaque red and green solid lines in Fig. \ref{fig6:theory}b(ii)).
Furthermore, comparable $R_d$ and $h$ allows the top and bottom walls to alter the rheotactic dynamics by significantly reducing $\bar{v}_r$ (see Eqs. (\ref{eq_vxdot}) and (\ref{eq_vydot})). 
We can conclude that while it is theoretically possible to model the oscillatory rheotaxis of a microswimmer by ignoring the aforementioned finite size effects, these observations will be aphysical, and inconsistent with the experimentally observed variations in $\bar{\Omega}$ and $\bar{v}_r$. 

The variation of $\bar{v}_r$ along the oscillatory trajectory obtained using Eqs. (\ref{eq_vxdot})-(\ref{eq_psidot}) (top half of Fig. \ref{fig6:theory}c) is also similar to the experimentally observed variation (Fig. \ref{fig1:rheotaxis_traj}e, g).
For a particular $\bar{u}_f^a$, as the active droplet approaches a side wall, the high $\bar{\Omega}$ (Fig. \ref{fig6:theory}e) due to its finite size (consequence of the source dipole) results in strong reorientation away from the wall.
This weakens the effect of the pusher-like droplet's attraction towards the approaching wall (the consequence of the force dipole).
Consequently, the increase in $\bar{v}_x$ (bottom half of Fig. \ref{fig6:theory}c) is weaker compared to the decrease in $\bar{v}_y$ (Fig. \ref{fig6:theory}d) in the direction of the approaching wall. 
Such relative variations of $\bar{v}_x$ and $\bar{v}_y$, and the enhanced magnitude of $\bar{\Omega}$, manifest in the sharp reductions in $\bar{v}_r$ (Fig. \ref{fig1:rheotaxis_traj}g) and $|\Psi|$ (Fig. \ref{fig3:orientation}a, b) during the approach towards a side wall.
However, in the immediate vicinity of a side wall, as $\bar{v}_y$ and $\Psi$ approach zero, the wall-induced sliding of the pusher-like droplet momentarily prevails resulting in a sharp increase in $\bar{v}_x$ (bottom half of Fig. \ref{fig6:theory}c).
This culminates in the high acceleration of the droplet adjacent to the wall (Fig. \ref{fig1:rheotaxis_traj}g).
Eventually, the strong $\bar{\Omega}$ orients the droplet towards the the opposite side wall with increasing $|\Psi|$ and $\bar{v}_y$ (Fig. \ref{fig6:theory}d).
However, during this part of the journey, the consequential reduction in $\bar{v}_x$ (bottom half of Fig. \ref{fig6:theory}c) dominates, resulting in a gradual decrease in $\bar{v}_r$ (Fig. \ref{fig1:rheotaxis_traj}g).
The aforementioned dynamics repeats itself once the droplet crosses the microchannel centerline, and approaches the opposite side wall.

Finally, Eqs. (\ref{eq_vxdot})-(\ref{eq_psidot}) also describe the downstream drift of the active droplet for $0.53<\bar{u}_f^a<1$ (compare Fig. \ref{fig6:theory}f and Fig. \ref{fig2:drifting_traj}).
The hydrodynamic model satisfactorily recreates the two main features of the swinging downstream drift- the active droplet always remains oriented upstream, and the wavelength of the swinging trajectory increases with increasing $\bar{u}_f^a$ (Fig. \ref{fig6:theory}f).
The consistent inclusion of the full hydrodynamic interaction of the finite-sized microswimmer with all the confinig walls, while estimating $\bar{v}_r$ and $\dot{\Psi}$, results in the recreation of the downstream drift for $\bar{u}_f^a<1$.
This is because now $\bar{v}_x$ becomes negative for a relatively smaller value of $\bar{u}_f^a$ due to the presence of the wall-induced hydrodynamic effects.
We also conclude that the theoretical model can be used to address the hydrodynamic trapping of the active droplet (Fig. \ref{fig1:rheotaxis_traj}h) by tuning $\bar{u}_f^a$ such that $\bar{\Omega}$ is strong and simultaneously $\bar{v}_x$ varies about $0$.
\section*{Discussion}
In this study, we have investigated in experiment and theory a problem common in nature and artificial microswimmers, e.g. lab-on-a-chip applications: robust positive rheotaxis in microchannels under external flow. 
We first document a novel oscillatory upstream rheotaxis in microchannels for artificial swimmers (active droplets).
Although this type of oscillatory rheotaxis has been observed for bioswimmers like \textit{E. coli} \cite{mathijssen2019oscillatory} and \textit{T. brucei} \cite{uppaluri2012_flow}, it has not been evidenced before for artificial microswimmers.
We demonstrate that the oscillatory rheotaxis of the active droplets can be physically captured by following a purely hydrodynamic approach of mapping our droplets on an ideal squirmer with a finite size, and then consistently accounting for its hydrodynamic interaction with the confining walls and the imposed shear flow.
We achieve good quantitative agreement between experiment and theory even under the approximations of ignoring the chemical interactions and truncating the hydrodynamic images systems to the leading order for the walls of the microchannel.
Using phase portraits, we illustrate that this model features a robust limit cycle, as also demonstrated by the striking regularity and reproducibility of our recorded trajectories. 
It will be interesting to investigate in future studies whether the inclusion of both phoretic and hydrodynamic interactions for the model microswimmer in the confinement, as done very recently \cite{choudhary2021self}, can bridge the remaining differences between experiment and our somewhat idealized model for the rheotaxis of artificial microswimmers - notably, the relative overestimation (underestimation) of the oscillation wavelength (angular velocity) during rheotaxis.
We strongly believe that the understanding of this new rheotactic behaviour of artificial microswimmers, including the possibility of hydrodynamically trapping these, will open up wide possibilities in conceptualizing and designing practical micro-robotic applications like targeted drug delivery.  

\section*{Methods}
\subsection*{Fabrication of micro-confinements}
For the production of CB15 droplets, we
fabricated microfluidic chips in-house, using
soft lithography techniques~\cite{qin2010_soft}, as follows: CAD photomasks were printed onto an emulsion
film in high-resolution (128 000 dpi) by a commercial
supplier (JD Photo-Tools). Next, the photoresist SU-8 3025
(MicroChem) was spin-coated onto a silicon
wafer (Si-Mat). A negative mold was
cured in the SU-8 by UV light
exposure through the photomask.\\
We then poured a poly(dimethyl siloxane) (PDMS),  mixture (SYLGARD 184 Silicone Elastomer Kit, DowCorning) with 10:1 weight ratio of base
to cross-linker over the wafer and baked it for 2 hours at 75 $^\circ$C.
For the droplet generation we used a glass slide as a coverslip for the channel, whereas for the microchannels used in the experiments we coated the glass with PDMS to have similar boundary conditions at all channel walls. To create an even PDMS surface, we placed the glass slide with liquid PDMS on top of it in a spin coater (Laurell, WS-650HZB-23NPPB) for 45 s at 500 rotations/min. Afterwards we baked it for 2 hours at 75 $^\circ$C.
Finally both channel structure and glass slide were treeated with a partial pressure air plasma (Pico P100-8; Diener Electronic GmbH+Co. KG) for 30 s and then pressed together, bonding the two surfaces.\\
\subsection*{Production of monodisperse CB15 droplets}
The walls of these microfluidic chips were chemically treated to hydrophilize the channels where aqueous surfactant solution flowed. We followed the recipe of Petit et al.~\cite{petit2016_vesicles-on-a-chip}: First, the channel walls were oxidized by a 1:1 mixture of hydrogen peroxide solution (H2O2 at 30 wt\%, Sigma-Aldrich) and hydrochloric acid (HCl at 37 wt\%,
Sigma-Aldrich). The mixture was flushed through the channels
for 2 minutes. Then, the channel was rinsed by flushing
double distilled water for 30 s. Next, a 5 wt\% solution of
the positive polyelectrolyte poly(diallyldimethylammonium
chloride) (PDADMAC, Sigma-Aldrich) was flushed for 2 minutes through the oxidized channel. The PDADMAC
binds to the activated channel walls by ionic interactions.
Finally, a 2 wt\% solution of the negative polyelectrolyte
poly(sodium 4-styrenesulfonate) (PSS, Sigma-Aldrich) was
flushed for 2 minutes.\\
Once the chips were treated, we mounted syringes of oil and
0.1 wt\% aqueous TTAB solution to a microprecision
syringe pump (NEM-B101-02B; Cetoni GmbH), connected
them to the two inlets of the microfluidic chip via Teflon
tubing, and tuned the flow speed
through the chip to reach the desired droplet diameter.
Once the production was monodisperse and at a steady state, droplets were
collected in a bath of 0.1 wt\% TTAB solution.\\
\subsection*{Experimental protocol for visualizing the rheotactic behaviour of active droplets}
To observe the activity of the droplets, we put them into a 7.5 w\% aqueous TTAB solution. To visualize the flow, and for subsequent PIV analysis, tracer particles (FluoSpheres\textsuperscript{TM} carboxylate, 0.5 $\mu$m, red (580/605), Life Technologies Corporation) were added to the aqueous solution.\\
\begin{figure}
\centering
\includegraphics[width=0.5\textwidth]{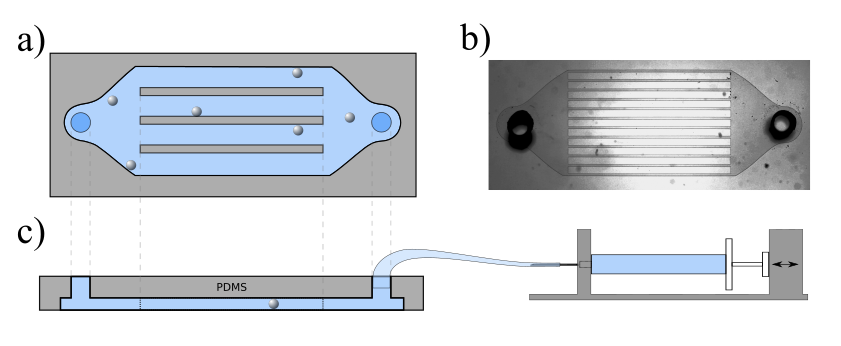}
 \caption{\label{fig7:channel_sketch} Sketch of the channel setup. a) Top view of the used channel structure, b) micrograph of the channel structure, c) side view of the PDMS structure with connected syringe pump.}
\end{figure}
We observed the droplet motion in a quasi-two-dimensional channel structure with several parallel channels (Fig. \ref{fig7:channel_sketch}), each of height $h = \SI{52}{\um}$ and width of w = 100$\mu$m. The flow inside the channels was controlled by a syringe pump connected via a Teflon tubing to the inlet of the channel structure. The outlet was left open and unconnected. We used an inverted bright-field microscope (Olympus IX-81) at a magnification of 20$\times$ for observation, and recorded videos at a frame rate of 25 frames per second with a Canon (EOS 600D) digital camera (1920$\times$1080 px). Pixel-to-micrometre conversions were calibrated by imaging microstructures with known dimensions, including micrometre stages and channel structures.
\subsection*{Post-processing of experimental data}
The droplet coordinates in each frame were extracted from video frames using software written in-house in Python/openCV\cite{bradski2000_opencv}, using a sequence of constant background correction, threshold binarization, blob detection by contour analysis, and minimum
enclosing circle fits. Swimming trajectories were obtained via a frame-by-frame nearest-neighbor analysis\cite{crocker1996_methods}.\\
We analyzed the flow in the aqueous swimming medium using the open source MATLAB package PIVlab 
\cite{thielicke2014_pivlab}, based on  images extracted from videomicroscopy data with the open source software ffmpeg, and without  any further preprocessing. For the calibration distance (inside PIVlab) the channel width was used with the defined width of 100$\mu$m. For the analysis, a region of interest is chosen over the channel cross section and of around 60 $\mu$m along the channel. Within PIVlab, the interrogation area was set specifically (Pass 1: 64, Pass 2: 32, Pass 3: 16, Pass 4: 8), with default settings kept otherwise. \\
Further calculations of the droplet speed, mean flow speed, etc. and plotting the figures were done with Python using mainly the libraries numpy and matplotlib.\\
\subsection*{Visualization of the chemical (filled micelle) trail during rheotaxis}
We used fluorescent microscopy for direct imaging of the chemical trail secreted by the droplet~\cite{hokmabad2021emergence, hokmabad2020quantitative}. We doped the oil phase with the fluorescent dye NileRed (Thermo Fisher Scientific). The dye molecules co-migrate with the oil phase into the filled micelles shed by the droplet. Experiments were observed on an Olympus IX73 microscope with a filter cube (Excitation filter  ET560/40x, Beam splitter 585 LP and Emissions filter ET630/75m, all by Chroma Technologies). Images were recorded using a 4 MegaPixels Grasshopper camera at 4 frames per second and $4\times$ magnification. Recorded images were analyzed, using MATLAB, to extract the red light intensity profiles ($I(\theta)$) around the droplet close to the interface. We then identified the angular position of the maximum intensity point $[I(\theta)]_{max}$. The line from this point to the center of the droplet determines the orientation of the chemical field and in turn yields the `chemical angle' $\Psi_c$.

\begin{acknowledgments}
CCM, BVH and RD acknowledge funding from the DFG SPP 1726 ``Microswimmers'', 
CCM and CJ from the BMBF/MPG MaxSynBio initiative. RD also acknowledges support from IIT Hyderabad.
\end{acknowledgments}
%
%


\begin{thebibliography}{38}%
\makeatletter
\providecommand \@ifxundefined [1]{%
 \@ifx{#1\undefined}
}%
\providecommand \@ifnum [1]{%
 \ifnum #1\expandafter \@firstoftwo
 \else \expandafter \@secondoftwo
 \fi
}%
\providecommand \@ifx [1]{%
 \ifx #1\expandafter \@firstoftwo
 \else \expandafter \@secondoftwo
 \fi
}%
\providecommand \natexlab [1]{#1}%
\providecommand \enquote  [1]{``#1''}%
\providecommand \bibnamefont  [1]{#1}%
\providecommand \bibfnamefont [1]{#1}%
\providecommand \citenamefont [1]{#1}%
\providecommand \href@noop [0]{\@secondoftwo}%
\providecommand \href [0]{\begingroup \@sanitize@url \@href}%
\providecommand \@href[1]{\@@startlink{#1}\@@href}%
\providecommand \@@href[1]{\endgroup#1\@@endlink}%
\providecommand \@sanitize@url [0]{\catcode `\\12\catcode `\$12\catcode
  `\&12\catcode `\#12\catcode `\^12\catcode `\_12\catcode `\%12\relax}%
\providecommand \@@startlink[1]{}%
\providecommand \@@endlink[0]{}%
\providecommand \url  [0]{\begingroup\@sanitize@url \@url }%
\providecommand \@url [1]{\endgroup\@href {#1}{\urlprefix }}%
\providecommand \urlprefix  [0]{URL }%
\providecommand \Eprint [0]{\href }%
\providecommand \doibase [0]{http://dx.doi.org/}%
\providecommand \selectlanguage [0]{\@gobble}%
\providecommand \bibinfo  [0]{\@secondoftwo}%
\providecommand \bibfield  [0]{\@secondoftwo}%
\providecommand \translation [1]{[#1]}%
\providecommand \BibitemOpen [0]{}%
\providecommand \bibitemStop [0]{}%
\providecommand \bibitemNoStop [0]{.\EOS\space}%
\providecommand \EOS [0]{\spacefactor3000\relax}%
\providecommand \BibitemShut  [1]{\csname bibitem#1\endcsname}%
\let\auto@bib@innerbib\@empty
\bibitem [{\citenamefont {Kantsler}\ \emph {et~al.}(2014)\citenamefont
  {Kantsler}, \citenamefont {Dunkel}, \citenamefont {Blayney},\ and\
  \citenamefont {Goldstein}}]{kantsler2014rheotaxis}%
  \BibitemOpen
  \bibfield  {author} {\bibinfo {author} {\bibfnamefont {V.}~\bibnamefont
  {Kantsler}}, \bibinfo {author} {\bibfnamefont {J.}~\bibnamefont {Dunkel}},
  \bibinfo {author} {\bibfnamefont {M.}~\bibnamefont {Blayney}}, \ and\
  \bibinfo {author} {\bibfnamefont {R.~E.}\ \bibnamefont {Goldstein}},\
  }\bibfield  {title} {\enquote {\bibinfo {title} {Rheotaxis facilitates
  upstream navigation of mammalian sperm cells},}\ }\href@noop {} {\bibfield
  {journal} {\bibinfo  {journal} {Elife}\ }\textbf {\bibinfo {volume} {3}},\
  \bibinfo {pages} {e02403} (\bibinfo {year} {2014})}\BibitemShut {NoStop}%
\bibitem [{\citenamefont {Lane}\ \emph {et~al.}(2005)\citenamefont {Lane},
  \citenamefont {Lockatell}, \citenamefont {Monterosso}, \citenamefont
  {Lamphier}, \citenamefont {Weinert}, \citenamefont {Hebel}, \citenamefont
  {Johnson},\ and\ \citenamefont {Mobley}}]{lane2005role}%
  \BibitemOpen
  \bibfield  {author} {\bibinfo {author} {\bibfnamefont {M.~C.}\ \bibnamefont
  {Lane}}, \bibinfo {author} {\bibfnamefont {V.}~\bibnamefont {Lockatell}},
  \bibinfo {author} {\bibfnamefont {G.}~\bibnamefont {Monterosso}}, \bibinfo
  {author} {\bibfnamefont {D.}~\bibnamefont {Lamphier}}, \bibinfo {author}
  {\bibfnamefont {J.}~\bibnamefont {Weinert}}, \bibinfo {author} {\bibfnamefont
  {J.~R.}\ \bibnamefont {Hebel}}, \bibinfo {author} {\bibfnamefont {D.~E.}\
  \bibnamefont {Johnson}}, \ and\ \bibinfo {author} {\bibfnamefont {H.~L.}\
  \bibnamefont {Mobley}},\ }\bibfield  {title} {\enquote {\bibinfo {title}
  {Role of motility in the colonization of uropathogenic escherichia coli in
  the urinary tract},}\ }\href@noop {} {\bibfield  {journal} {\bibinfo
  {journal} {Infection and immunity}\ }\textbf {\bibinfo {volume} {73}},\
  \bibinfo {pages} {7644--7656} (\bibinfo {year} {2005})}\BibitemShut {NoStop}%
\bibitem [{\citenamefont {Figueroa-Morales}\ \emph {et~al.}(2020)\citenamefont
  {Figueroa-Morales}, \citenamefont {Rivera}, \citenamefont {Soto},
  \citenamefont {Lindner}, \citenamefont {Altshuler},\ and\ \citenamefont
  {Cl{\'e}ment}}]{figueroa2020coli}%
  \BibitemOpen
  \bibfield  {author} {\bibinfo {author} {\bibfnamefont {N.}~\bibnamefont
  {Figueroa-Morales}}, \bibinfo {author} {\bibfnamefont {A.}~\bibnamefont
  {Rivera}}, \bibinfo {author} {\bibfnamefont {R.}~\bibnamefont {Soto}},
  \bibinfo {author} {\bibfnamefont {A.}~\bibnamefont {Lindner}}, \bibinfo
  {author} {\bibfnamefont {E.}~\bibnamefont {Altshuler}}, \ and\ \bibinfo
  {author} {\bibfnamefont {{\'E}.}~\bibnamefont {Cl{\'e}ment}},\ }\bibfield
  {title} {\enquote {\bibinfo {title} {E. coli “super-contaminates” narrow
  ducts fostered by broad run-time distribution},}\ }\href@noop {} {\bibfield
  {journal} {\bibinfo  {journal} {Science advances}\ }\textbf {\bibinfo
  {volume} {6}},\ \bibinfo {pages} {eaay0155} (\bibinfo {year}
  {2020})}\BibitemShut {NoStop}%
\bibitem [{\citenamefont {Bretherton}\ and\ \citenamefont
  {Rothschild}(1961)}]{bretherton1961rheotaxis}%
  \BibitemOpen
  \bibfield  {author} {\bibinfo {author} {\bibfnamefont {F.}~\bibnamefont
  {Bretherton}}\ and\ \bibinfo {author} {\bibfnamefont {N.~M.~V.}\ \bibnamefont
  {Rothschild}},\ }\bibfield  {title} {\enquote {\bibinfo {title} {Rheotaxis of
  spermatozoa},}\ }\href@noop {} {\bibfield  {journal} {\bibinfo  {journal}
  {Proceedings of the Royal Society of London. Series B. Biological Sciences}\
  }\textbf {\bibinfo {volume} {153}},\ \bibinfo {pages} {490--502} (\bibinfo
  {year} {1961})}\BibitemShut {NoStop}%
\bibitem [{\citenamefont {Kaya}\ and\ \citenamefont
  {Koser}(2012)}]{kaya2012direct}%
  \BibitemOpen
  \bibfield  {author} {\bibinfo {author} {\bibfnamefont {T.}~\bibnamefont
  {Kaya}}\ and\ \bibinfo {author} {\bibfnamefont {H.}~\bibnamefont {Koser}},\
  }\bibfield  {title} {\enquote {\bibinfo {title} {Direct upstream motility in
  escherichia coli},}\ }\href@noop {} {\bibfield  {journal} {\bibinfo
  {journal} {Biophysical journal}\ }\textbf {\bibinfo {volume} {102}},\
  \bibinfo {pages} {1514--1523} (\bibinfo {year} {2012})}\BibitemShut {NoStop}%
\bibitem [{\citenamefont {Uppaluri}\ \emph {et~al.}(2012)\citenamefont
  {Uppaluri}, \citenamefont {Heddergott}, \citenamefont {Stellamanns},
  \citenamefont {Herminghaus}, \citenamefont {Z{\"o}ttl}, \citenamefont
  {Stark}, \citenamefont {Engstler},\ and\ \citenamefont
  {Pfohl}}]{uppaluri2012_flow}%
  \BibitemOpen
  \bibfield  {author} {\bibinfo {author} {\bibfnamefont {S.}~\bibnamefont
  {Uppaluri}}, \bibinfo {author} {\bibfnamefont {N.}~\bibnamefont
  {Heddergott}}, \bibinfo {author} {\bibfnamefont {E.}~\bibnamefont
  {Stellamanns}}, \bibinfo {author} {\bibfnamefont {S.}~\bibnamefont
  {Herminghaus}}, \bibinfo {author} {\bibfnamefont {A.}~\bibnamefont
  {Z{\"o}ttl}}, \bibinfo {author} {\bibfnamefont {H.}~\bibnamefont {Stark}},
  \bibinfo {author} {\bibfnamefont {M.}~\bibnamefont {Engstler}}, \ and\
  \bibinfo {author} {\bibfnamefont {T.}~\bibnamefont {Pfohl}},\ }\bibfield
  {title} {\enquote {\bibinfo {title} {Flow loading induces oscillatory
  trajectories in a bloodstream parasite},}\ }\href@noop {} {\bibfield
  {journal} {\bibinfo  {journal} {Biophysical Journal}\ }\textbf {\bibinfo
  {volume} {103}},\ \bibinfo {pages} {1162--1169} (\bibinfo {year}
  {2012})}\BibitemShut {NoStop}%
\bibitem [{\citenamefont {Fu}\ \emph {et~al.}(2012)\citenamefont {Fu},
  \citenamefont {Powers}, \citenamefont {Stocker} \emph
  {et~al.}}]{fu2012bacterial}%
  \BibitemOpen
  \bibfield  {author} {\bibinfo {author} {\bibfnamefont {H.~C.}\ \bibnamefont
  {Fu}}, \bibinfo {author} {\bibfnamefont {T.~R.}\ \bibnamefont {Powers}},
  \bibinfo {author} {\bibfnamefont {R.}~\bibnamefont {Stocker}},  \emph
  {et~al.},\ }\bibfield  {title} {\enquote {\bibinfo {title} {Bacterial
  rheotaxis},}\ }\href@noop {} {\bibfield  {journal} {\bibinfo  {journal}
  {Proceedings of the National Academy of Sciences}\ }\textbf {\bibinfo
  {volume} {109}},\ \bibinfo {pages} {4780--4785} (\bibinfo {year}
  {2012})}\BibitemShut {NoStop}%
\bibitem [{\citenamefont {Figueroa-Morales}\ \emph {et~al.}(2015)\citenamefont
  {Figueroa-Morales}, \citenamefont {Mino}, \citenamefont {Rivera},
  \citenamefont {Caballero}, \citenamefont {Cl{\'e}ment}, \citenamefont
  {Altshuler},\ and\ \citenamefont {Lindner}}]{figueroa2015living}%
  \BibitemOpen
  \bibfield  {author} {\bibinfo {author} {\bibfnamefont {N.}~\bibnamefont
  {Figueroa-Morales}}, \bibinfo {author} {\bibfnamefont {G.~L.}\ \bibnamefont
  {Mino}}, \bibinfo {author} {\bibfnamefont {A.}~\bibnamefont {Rivera}},
  \bibinfo {author} {\bibfnamefont {R.}~\bibnamefont {Caballero}}, \bibinfo
  {author} {\bibfnamefont {E.}~\bibnamefont {Cl{\'e}ment}}, \bibinfo {author}
  {\bibfnamefont {E.}~\bibnamefont {Altshuler}}, \ and\ \bibinfo {author}
  {\bibfnamefont {A.}~\bibnamefont {Lindner}},\ }\bibfield  {title} {\enquote
  {\bibinfo {title} {Living on the edge: transfer and traffic of e. coli in a
  confined flow},}\ }\href@noop {} {\bibfield  {journal} {\bibinfo  {journal}
  {Soft matter}\ }\textbf {\bibinfo {volume} {11}},\ \bibinfo {pages}
  {6284--6293} (\bibinfo {year} {2015})}\BibitemShut {NoStop}%
\bibitem [{\citenamefont {Junot}\ \emph {et~al.}(2019)\citenamefont {Junot},
  \citenamefont {Figueroa-Morales}, \citenamefont {Darnige}, \citenamefont
  {Lindner}, \citenamefont {Soto}, \citenamefont {Auradou},\ and\ \citenamefont
  {Cl{\'e}ment}}]{junot2019swimming}%
  \BibitemOpen
  \bibfield  {author} {\bibinfo {author} {\bibfnamefont {G.}~\bibnamefont
  {Junot}}, \bibinfo {author} {\bibfnamefont {N.}~\bibnamefont
  {Figueroa-Morales}}, \bibinfo {author} {\bibfnamefont {T.}~\bibnamefont
  {Darnige}}, \bibinfo {author} {\bibfnamefont {A.}~\bibnamefont {Lindner}},
  \bibinfo {author} {\bibfnamefont {R.}~\bibnamefont {Soto}}, \bibinfo {author}
  {\bibfnamefont {H.}~\bibnamefont {Auradou}}, \ and\ \bibinfo {author}
  {\bibfnamefont {E.}~\bibnamefont {Cl{\'e}ment}},\ }\bibfield  {title}
  {\enquote {\bibinfo {title} {Swimming bacteria in poiseuille flow: The quest
  for active bretherton-jeffery trajectories},}\ }\href@noop {} {\bibfield
  {journal} {\bibinfo  {journal} {EPL (Europhysics Letters)}\ }\textbf
  {\bibinfo {volume} {126}},\ \bibinfo {pages} {44003} (\bibinfo {year}
  {2019})}\BibitemShut {NoStop}%
\bibitem [{\citenamefont {Mathijssen}\ \emph {et~al.}(2019)\citenamefont
  {Mathijssen}, \citenamefont {Figueroa-Morales}, \citenamefont {Junot},
  \citenamefont {Cl{\'e}ment}, \citenamefont {Lindner},\ and\ \citenamefont
  {Z{\"o}ttl}}]{mathijssen2019oscillatory}%
  \BibitemOpen
  \bibfield  {author} {\bibinfo {author} {\bibfnamefont {A.~J.}\ \bibnamefont
  {Mathijssen}}, \bibinfo {author} {\bibfnamefont {N.}~\bibnamefont
  {Figueroa-Morales}}, \bibinfo {author} {\bibfnamefont {G.}~\bibnamefont
  {Junot}}, \bibinfo {author} {\bibfnamefont {{\'E}.}~\bibnamefont
  {Cl{\'e}ment}}, \bibinfo {author} {\bibfnamefont {A.}~\bibnamefont
  {Lindner}}, \ and\ \bibinfo {author} {\bibfnamefont {A.}~\bibnamefont
  {Z{\"o}ttl}},\ }\bibfield  {title} {\enquote {\bibinfo {title} {Oscillatory
  surface rheotaxis of swimming e. coli bacteria},}\ }\href@noop {} {\bibfield
  {journal} {\bibinfo  {journal} {Nature communications}\ }\textbf {\bibinfo
  {volume} {10}},\ \bibinfo {pages} {1--12} (\bibinfo {year}
  {2019})}\BibitemShut {NoStop}%
\bibitem [{\citenamefont {Jing}\ \emph {et~al.}(2020)\citenamefont {Jing},
  \citenamefont {Z{\"o}ttl}, \citenamefont {Cl{\'e}ment},\ and\ \citenamefont
  {Lindner}}]{jing2020chirality}%
  \BibitemOpen
  \bibfield  {author} {\bibinfo {author} {\bibfnamefont {G.}~\bibnamefont
  {Jing}}, \bibinfo {author} {\bibfnamefont {A.}~\bibnamefont {Z{\"o}ttl}},
  \bibinfo {author} {\bibfnamefont {{\'E}.}~\bibnamefont {Cl{\'e}ment}}, \ and\
  \bibinfo {author} {\bibfnamefont {A.}~\bibnamefont {Lindner}},\ }\bibfield
  {title} {\enquote {\bibinfo {title} {Chirality-induced bacterial rheotaxis in
  bulk shear flows},}\ }\href@noop {} {\bibfield  {journal} {\bibinfo
  {journal} {Science advances}\ }\textbf {\bibinfo {volume} {6}},\ \bibinfo
  {pages} {eabb2012} (\bibinfo {year} {2020})}\BibitemShut {NoStop}%
\bibitem [{\citenamefont {Soler}\ and\ \citenamefont
  {S{\'a}nchez}(2014)}]{soler2014catalytic}%
  \BibitemOpen
  \bibfield  {author} {\bibinfo {author} {\bibfnamefont {L.}~\bibnamefont
  {Soler}}\ and\ \bibinfo {author} {\bibfnamefont {S.}~\bibnamefont
  {S{\'a}nchez}},\ }\bibfield  {title} {\enquote {\bibinfo {title} {Catalytic
  nanomotors for environmental monitoring and water remediation},}\ }\href@noop
  {} {\bibfield  {journal} {\bibinfo  {journal} {Nanoscale}\ }\textbf {\bibinfo
  {volume} {6}},\ \bibinfo {pages} {7175--7182} (\bibinfo {year}
  {2014})}\BibitemShut {NoStop}%
\bibitem [{\citenamefont {Gao}\ and\ \citenamefont
  {Wang}(2014)}]{gao2014environmental}%
  \BibitemOpen
  \bibfield  {author} {\bibinfo {author} {\bibfnamefont {W.}~\bibnamefont
  {Gao}}\ and\ \bibinfo {author} {\bibfnamefont {J.}~\bibnamefont {Wang}},\
  }\bibfield  {title} {\enquote {\bibinfo {title} {The environmental impact of
  micro/nanomachines: a review},}\ }\href@noop {} {\bibfield  {journal}
  {\bibinfo  {journal} {Acs Nano}\ }\textbf {\bibinfo {volume} {8}},\ \bibinfo
  {pages} {3170--3180} (\bibinfo {year} {2014})}\BibitemShut {NoStop}%
\bibitem [{\citenamefont {Ren}\ \emph {et~al.}(2017)\citenamefont {Ren},
  \citenamefont {Zhou}, \citenamefont {Mao}, \citenamefont {Xu}, \citenamefont
  {Huang},\ and\ \citenamefont {Mallouk}}]{ren2017rheotaxis}%
  \BibitemOpen
  \bibfield  {author} {\bibinfo {author} {\bibfnamefont {L.}~\bibnamefont
  {Ren}}, \bibinfo {author} {\bibfnamefont {D.}~\bibnamefont {Zhou}}, \bibinfo
  {author} {\bibfnamefont {Z.}~\bibnamefont {Mao}}, \bibinfo {author}
  {\bibfnamefont {P.}~\bibnamefont {Xu}}, \bibinfo {author} {\bibfnamefont
  {T.~J.}\ \bibnamefont {Huang}}, \ and\ \bibinfo {author} {\bibfnamefont
  {T.~E.}\ \bibnamefont {Mallouk}},\ }\bibfield  {title} {\enquote {\bibinfo
  {title} {Rheotaxis of bimetallic micromotors driven by chemical--acoustic
  hybrid power},}\ }\href@noop {} {\bibfield  {journal} {\bibinfo  {journal}
  {ACS nano}\ }\textbf {\bibinfo {volume} {11}},\ \bibinfo {pages}
  {10591--10598} (\bibinfo {year} {2017})}\BibitemShut {NoStop}%
\bibitem [{\citenamefont {Katuri}\ \emph {et~al.}(2018)\citenamefont {Katuri},
  \citenamefont {Uspal}, \citenamefont {Simmchen}, \citenamefont
  {Miguel-L{\'o}pez},\ and\ \citenamefont {S{\'a}nchez}}]{katuri2018cross}%
  \BibitemOpen
  \bibfield  {author} {\bibinfo {author} {\bibfnamefont {J.}~\bibnamefont
  {Katuri}}, \bibinfo {author} {\bibfnamefont {W.~E.}\ \bibnamefont {Uspal}},
  \bibinfo {author} {\bibfnamefont {J.}~\bibnamefont {Simmchen}}, \bibinfo
  {author} {\bibfnamefont {A.}~\bibnamefont {Miguel-L{\'o}pez}}, \ and\
  \bibinfo {author} {\bibfnamefont {S.}~\bibnamefont {S{\'a}nchez}},\
  }\bibfield  {title} {\enquote {\bibinfo {title} {Cross-stream migration of
  active particles},}\ }\href@noop {} {\bibfield  {journal} {\bibinfo
  {journal} {Science advances}\ }\textbf {\bibinfo {volume} {4}},\ \bibinfo
  {pages} {eaao1755} (\bibinfo {year} {2018})}\BibitemShut {NoStop}%
\bibitem [{\citenamefont {Uspal}\ \emph {et~al.}(2015)\citenamefont {Uspal},
  \citenamefont {Popescu}, \citenamefont {Dietrich},\ and\ \citenamefont
  {Tasinkevych}}]{uspal2015rheotaxis}%
  \BibitemOpen
  \bibfield  {author} {\bibinfo {author} {\bibfnamefont {W.}~\bibnamefont
  {Uspal}}, \bibinfo {author} {\bibfnamefont {M.~N.}\ \bibnamefont {Popescu}},
  \bibinfo {author} {\bibfnamefont {S.}~\bibnamefont {Dietrich}}, \ and\
  \bibinfo {author} {\bibfnamefont {M.}~\bibnamefont {Tasinkevych}},\
  }\bibfield  {title} {\enquote {\bibinfo {title} {Rheotaxis of spherical
  active particles near a planar wall},}\ }\href@noop {} {\bibfield  {journal}
  {\bibinfo  {journal} {Soft Matter}\ }\textbf {\bibinfo {volume} {11}},\
  \bibinfo {pages} {6613--6632} (\bibinfo {year} {2015})}\BibitemShut {NoStop}%
\bibitem [{\citenamefont {Brosseau}\ \emph {et~al.}(2019)\citenamefont
  {Brosseau}, \citenamefont {Usabiaga}, \citenamefont {Lushi}, \citenamefont
  {Wu}, \citenamefont {Ristroph}, \citenamefont {Zhang}, \citenamefont {Ward},\
  and\ \citenamefont {Shelley}}]{brosseau2019relating}%
  \BibitemOpen
  \bibfield  {author} {\bibinfo {author} {\bibfnamefont {Q.}~\bibnamefont
  {Brosseau}}, \bibinfo {author} {\bibfnamefont {F.~B.}\ \bibnamefont
  {Usabiaga}}, \bibinfo {author} {\bibfnamefont {E.}~\bibnamefont {Lushi}},
  \bibinfo {author} {\bibfnamefont {Y.}~\bibnamefont {Wu}}, \bibinfo {author}
  {\bibfnamefont {L.}~\bibnamefont {Ristroph}}, \bibinfo {author}
  {\bibfnamefont {J.}~\bibnamefont {Zhang}}, \bibinfo {author} {\bibfnamefont
  {M.}~\bibnamefont {Ward}}, \ and\ \bibinfo {author} {\bibfnamefont {M.~J.}\
  \bibnamefont {Shelley}},\ }\bibfield  {title} {\enquote {\bibinfo {title}
  {Relating rheotaxis and hydrodynamic actuation using asymmetric gold-platinum
  phoretic rods},}\ }\href@noop {} {\bibfield  {journal} {\bibinfo  {journal}
  {Physical review letters}\ }\textbf {\bibinfo {volume} {123}},\ \bibinfo
  {pages} {178004} (\bibinfo {year} {2019})}\BibitemShut {NoStop}%
\bibitem [{\citenamefont {Baker}\ \emph {et~al.}(2019)\citenamefont {Baker},
  \citenamefont {Kauffman}, \citenamefont {Laskar}, \citenamefont {Shklyaev},
  \citenamefont {Potomkin}, \citenamefont {Dominguez-Rubio}, \citenamefont
  {Shum}, \citenamefont {Cruz-Rivera}, \citenamefont {Aranson}, \citenamefont
  {Balazs} \emph {et~al.}}]{baker2019fight}%
  \BibitemOpen
  \bibfield  {author} {\bibinfo {author} {\bibfnamefont {R.}~\bibnamefont
  {Baker}}, \bibinfo {author} {\bibfnamefont {J.~E.}\ \bibnamefont {Kauffman}},
  \bibinfo {author} {\bibfnamefont {A.}~\bibnamefont {Laskar}}, \bibinfo
  {author} {\bibfnamefont {O.~E.}\ \bibnamefont {Shklyaev}}, \bibinfo {author}
  {\bibfnamefont {M.}~\bibnamefont {Potomkin}}, \bibinfo {author}
  {\bibfnamefont {L.}~\bibnamefont {Dominguez-Rubio}}, \bibinfo {author}
  {\bibfnamefont {H.}~\bibnamefont {Shum}}, \bibinfo {author} {\bibfnamefont
  {Y.}~\bibnamefont {Cruz-Rivera}}, \bibinfo {author} {\bibfnamefont {I.~S.}\
  \bibnamefont {Aranson}}, \bibinfo {author} {\bibfnamefont {A.~C.}\
  \bibnamefont {Balazs}},  \emph {et~al.},\ }\bibfield  {title} {\enquote
  {\bibinfo {title} {Fight the flow: the role of shear in artificial rheotaxis
  for individual and collective motion},}\ }\href@noop {} {\bibfield  {journal}
  {\bibinfo  {journal} {Nanoscale}\ }\textbf {\bibinfo {volume} {11}},\
  \bibinfo {pages} {10944--10951} (\bibinfo {year} {2019})}\BibitemShut
  {NoStop}%
\bibitem [{\citenamefont {Maass}\ \emph {et~al.}(2016)\citenamefont {Maass},
  \citenamefont {Kr{\"u}ger}, \citenamefont {Herminghaus},\ and\ \citenamefont
  {Bahr}}]{maass2016_swimming}%
  \BibitemOpen
  \bibfield  {author} {\bibinfo {author} {\bibfnamefont {C.~C.}\ \bibnamefont
  {Maass}}, \bibinfo {author} {\bibfnamefont {C.}~\bibnamefont {Kr{\"u}ger}},
  \bibinfo {author} {\bibfnamefont {S.}~\bibnamefont {Herminghaus}}, \ and\
  \bibinfo {author} {\bibfnamefont {C.}~\bibnamefont {Bahr}},\ }\bibfield
  {title} {\enquote {\bibinfo {title} {Swimming {{Droplets}}},}\ }\href@noop {}
  {\bibfield  {journal} {\bibinfo  {journal} {Annu. Rev. Condens. Matter
  Phys.}\ }\textbf {\bibinfo {volume} {7}},\ \bibinfo {pages} {171--193}
  (\bibinfo {year} {2016})}\BibitemShut {NoStop}%
\bibitem [{\citenamefont {Izri}\ \emph {et~al.}(2014)\citenamefont {Izri},
  \citenamefont {Van Der~Linden}, \citenamefont {Michelin},\ and\ \citenamefont
  {Dauchot}}]{izri2014self}%
  \BibitemOpen
  \bibfield  {author} {\bibinfo {author} {\bibfnamefont {Z.}~\bibnamefont
  {Izri}}, \bibinfo {author} {\bibfnamefont {M.~N.}\ \bibnamefont {Van
  Der~Linden}}, \bibinfo {author} {\bibfnamefont {S.}~\bibnamefont {Michelin}},
  \ and\ \bibinfo {author} {\bibfnamefont {O.}~\bibnamefont {Dauchot}},\
  }\bibfield  {title} {\enquote {\bibinfo {title} {Self-propulsion of pure
  water droplets by spontaneous marangoni-stress-driven motion},}\ }\href@noop
  {} {\bibfield  {journal} {\bibinfo  {journal} {Physical review letters}\
  }\textbf {\bibinfo {volume} {113}},\ \bibinfo {pages} {248302} (\bibinfo
  {year} {2014})}\BibitemShut {NoStop}%
\bibitem [{\citenamefont {Morozov}\ and\ \citenamefont
  {Michelin}(2019)}]{morozov2019nonlinear}%
  \BibitemOpen
  \bibfield  {author} {\bibinfo {author} {\bibfnamefont {M.}~\bibnamefont
  {Morozov}}\ and\ \bibinfo {author} {\bibfnamefont {S.}~\bibnamefont
  {Michelin}},\ }\bibfield  {title} {\enquote {\bibinfo {title} {Nonlinear
  dynamics of a chemically-active drop: From steady to chaotic
  self-propulsion},}\ }\href@noop {} {\bibfield  {journal} {\bibinfo  {journal}
  {The Journal of chemical physics}\ }\textbf {\bibinfo {volume} {150}},\
  \bibinfo {pages} {044110} (\bibinfo {year} {2019})}\BibitemShut {NoStop}%
\bibitem [{\citenamefont {Hokmabad}\ \emph {et~al.}(2021)\citenamefont
  {Hokmabad}, \citenamefont {Dey}, \citenamefont {Jalaal}, \citenamefont
  {Mohanty}, \citenamefont {Almukambetova}, \citenamefont {Baldwin},
  \citenamefont {Lohse},\ and\ \citenamefont {Maass}}]{hokmabad2021emergence}%
  \BibitemOpen
  \bibfield  {author} {\bibinfo {author} {\bibfnamefont {B.~V.}\ \bibnamefont
  {Hokmabad}}, \bibinfo {author} {\bibfnamefont {R.}~\bibnamefont {Dey}},
  \bibinfo {author} {\bibfnamefont {M.}~\bibnamefont {Jalaal}}, \bibinfo
  {author} {\bibfnamefont {D.}~\bibnamefont {Mohanty}}, \bibinfo {author}
  {\bibfnamefont {M.}~\bibnamefont {Almukambetova}}, \bibinfo {author}
  {\bibfnamefont {K.~A.}\ \bibnamefont {Baldwin}}, \bibinfo {author}
  {\bibfnamefont {D.}~\bibnamefont {Lohse}}, \ and\ \bibinfo {author}
  {\bibfnamefont {C.~C.}\ \bibnamefont {Maass}},\ }\bibfield  {title} {\enquote
  {\bibinfo {title} {Emergence of bimodal motility in active droplets},}\
  }\href@noop {} {\bibfield  {journal} {\bibinfo  {journal} {Physical Review
  X}\ }\textbf {\bibinfo {volume} {11}},\ \bibinfo {pages} {011043} (\bibinfo
  {year} {2021})}\BibitemShut {NoStop}%
\bibitem [{\citenamefont {Z{\"o}ttl}\ and\ \citenamefont
  {Stark}(2012)}]{zottl2012_nonlinear}%
  \BibitemOpen
  \bibfield  {author} {\bibinfo {author} {\bibfnamefont {A.}~\bibnamefont
  {Z{\"o}ttl}}\ and\ \bibinfo {author} {\bibfnamefont {H.}~\bibnamefont
  {Stark}},\ }\bibfield  {title} {\enquote {\bibinfo {title} {Nonlinear
  {{Dynamics}} of a {{Microswimmer}} in {{Poiseuille Flow}}},}\ }\href@noop {}
  {\bibfield  {journal} {\bibinfo  {journal} {Phys. Rev. Lett.}\ }\textbf
  {\bibinfo {volume} {108}},\ \bibinfo {pages} {218104} (\bibinfo {year}
  {2012})}\BibitemShut {NoStop}%
\bibitem [{\citenamefont {Stark}(2016)}]{stark2016swimming}%
  \BibitemOpen
  \bibfield  {author} {\bibinfo {author} {\bibfnamefont {H.}~\bibnamefont
  {Stark}},\ }\bibfield  {title} {\enquote {\bibinfo {title} {Swimming in
  external fields},}\ }\href@noop {} {\bibfield  {journal} {\bibinfo  {journal}
  {The European Physical Journal Special Topics}\ }\textbf {\bibinfo {volume}
  {225}},\ \bibinfo {pages} {2369--2387} (\bibinfo {year} {2016})}\BibitemShut
  {NoStop}%
\bibitem [{\citenamefont {Spagnolie}\ and\ \citenamefont
  {Lauga}(2012)}]{spagnolie2012_hydrodynamics}%
  \BibitemOpen
  \bibfield  {author} {\bibinfo {author} {\bibfnamefont {S.~E.}\ \bibnamefont
  {Spagnolie}}\ and\ \bibinfo {author} {\bibfnamefont {E.}~\bibnamefont
  {Lauga}},\ }\bibfield  {title} {\enquote {\bibinfo {title} {Hydrodynamics of
  self-propulsion near a boundary: Predictions and accuracy of far-field
  approximations},}\ }\href@noop {} {\bibfield  {journal} {\bibinfo  {journal}
  {Journal of Fluid Mechanics}\ }\textbf {\bibinfo {volume} {700}},\ \bibinfo
  {pages} {105--147} (\bibinfo {year} {2012})}\BibitemShut {NoStop}%
\bibitem [{\citenamefont {Bretherton}(1962)}]{bretherton1962motion}%
  \BibitemOpen
  \bibfield  {author} {\bibinfo {author} {\bibfnamefont {F.~P.}\ \bibnamefont
  {Bretherton}},\ }\bibfield  {title} {\enquote {\bibinfo {title} {The motion
  of rigid particles in a shear flow at low reynolds number},}\ }\href@noop {}
  {\bibfield  {journal} {\bibinfo  {journal} {Journal of Fluid Mechanics}\
  }\textbf {\bibinfo {volume} {14}},\ \bibinfo {pages} {284--304} (\bibinfo
  {year} {1962})}\BibitemShut {NoStop}%
\bibitem [{\citenamefont {de~Blois}\ \emph {et~al.}(2019)\citenamefont
  {de~Blois}, \citenamefont {Reyssat}, \citenamefont {Michelin},\ and\
  \citenamefont {Dauchot}}]{de2019flow}%
  \BibitemOpen
  \bibfield  {author} {\bibinfo {author} {\bibfnamefont {C.}~\bibnamefont
  {de~Blois}}, \bibinfo {author} {\bibfnamefont {M.}~\bibnamefont {Reyssat}},
  \bibinfo {author} {\bibfnamefont {S.}~\bibnamefont {Michelin}}, \ and\
  \bibinfo {author} {\bibfnamefont {O.}~\bibnamefont {Dauchot}},\ }\bibfield
  {title} {\enquote {\bibinfo {title} {Flow field around a confined active
  droplet},}\ }\href@noop {} {\bibfield  {journal} {\bibinfo  {journal}
  {Physical Review Fluids}\ }\textbf {\bibinfo {volume} {4}},\ \bibinfo {pages}
  {054001} (\bibinfo {year} {2019})}\BibitemShut {NoStop}%
\bibitem [{\citenamefont {Kuron}\ \emph {et~al.}(2019)\citenamefont {Kuron},
  \citenamefont {St{\"a}rk}, \citenamefont {Holm},\ and\ \citenamefont
  {De~Graaf}}]{kuron2019hydrodynamic}%
  \BibitemOpen
  \bibfield  {author} {\bibinfo {author} {\bibfnamefont {M.}~\bibnamefont
  {Kuron}}, \bibinfo {author} {\bibfnamefont {P.}~\bibnamefont {St{\"a}rk}},
  \bibinfo {author} {\bibfnamefont {C.}~\bibnamefont {Holm}}, \ and\ \bibinfo
  {author} {\bibfnamefont {J.}~\bibnamefont {De~Graaf}},\ }\bibfield  {title}
  {\enquote {\bibinfo {title} {Hydrodynamic mobility reversal of squirmers near
  flat and curved surfaces},}\ }\href@noop {} {\bibfield  {journal} {\bibinfo
  {journal} {Soft matter}\ }\textbf {\bibinfo {volume} {15}},\ \bibinfo {pages}
  {5908--5920} (\bibinfo {year} {2019})}\BibitemShut {NoStop}%
\bibitem [{\citenamefont {de~Graaf}\ \emph {et~al.}(2016)\citenamefont
  {de~Graaf}, \citenamefont {Mathijssen}, \citenamefont {Fabritius},
  \citenamefont {Menke}, \citenamefont {Holm},\ and\ \citenamefont
  {Shendruk}}]{de2016understanding}%
  \BibitemOpen
  \bibfield  {author} {\bibinfo {author} {\bibfnamefont {J.}~\bibnamefont
  {de~Graaf}}, \bibinfo {author} {\bibfnamefont {A.~J.}\ \bibnamefont
  {Mathijssen}}, \bibinfo {author} {\bibfnamefont {M.}~\bibnamefont
  {Fabritius}}, \bibinfo {author} {\bibfnamefont {H.}~\bibnamefont {Menke}},
  \bibinfo {author} {\bibfnamefont {C.}~\bibnamefont {Holm}}, \ and\ \bibinfo
  {author} {\bibfnamefont {T.~N.}\ \bibnamefont {Shendruk}},\ }\bibfield
  {title} {\enquote {\bibinfo {title} {Understanding the onset of oscillatory
  swimming in microchannels},}\ }\href@noop {} {\bibfield  {journal} {\bibinfo
  {journal} {Soft Matter}\ }\textbf {\bibinfo {volume} {12}},\ \bibinfo {pages}
  {4704--4708} (\bibinfo {year} {2016})}\BibitemShut {NoStop}%
\bibitem [{\citenamefont {Mathijssen}\ \emph {et~al.}(2016)\citenamefont
  {Mathijssen}, \citenamefont {Doostmohammadi}, \citenamefont {Yeomans},\ and\
  \citenamefont {Shendruk}}]{mathijssen2016hydrodynamics}%
  \BibitemOpen
  \bibfield  {author} {\bibinfo {author} {\bibfnamefont {A.~J.}\ \bibnamefont
  {Mathijssen}}, \bibinfo {author} {\bibfnamefont {A.}~\bibnamefont
  {Doostmohammadi}}, \bibinfo {author} {\bibfnamefont {J.~M.}\ \bibnamefont
  {Yeomans}}, \ and\ \bibinfo {author} {\bibfnamefont {T.~N.}\ \bibnamefont
  {Shendruk}},\ }\bibfield  {title} {\enquote {\bibinfo {title} {Hydrodynamics
  of micro-swimmers in films},}\ }\href@noop {} {\bibfield  {journal} {\bibinfo
   {journal} {Journal of Fluid Mechanics}\ }\textbf {\bibinfo {volume} {806}},\
  \bibinfo {pages} {35--70} (\bibinfo {year} {2016})}\BibitemShut {NoStop}%
\bibitem [{\citenamefont {Leal}(2007)}]{leal2007_advanced}%
  \BibitemOpen
  \bibfield  {author} {\bibinfo {author} {\bibfnamefont {L.~G.}\ \bibnamefont
  {Leal}},\ }\href@noop {} {\emph {\bibinfo {title} {Advanced Transport
  Phenomena: {{Fluid}} Mechanics and Convective Transport Processes}}},\
  Cambridge Series in Chemical Engineering\ (\bibinfo  {publisher} {{Cambridge
  University Press}},\ \bibinfo {address} {{Cambridge}},\ \bibinfo {year}
  {2007})\BibitemShut {NoStop}%
\bibitem [{\citenamefont {Choudhary}\ \emph {et~al.}(2021)\citenamefont
  {Choudhary}, \citenamefont {Chaithanya}, \citenamefont {Michelin},\ and\
  \citenamefont {Pushpavanam}}]{choudhary2021self}%
  \BibitemOpen
  \bibfield  {author} {\bibinfo {author} {\bibfnamefont {A.}~\bibnamefont
  {Choudhary}}, \bibinfo {author} {\bibfnamefont {K.}~\bibnamefont
  {Chaithanya}}, \bibinfo {author} {\bibfnamefont {S.}~\bibnamefont
  {Michelin}}, \ and\ \bibinfo {author} {\bibfnamefont {S.}~\bibnamefont
  {Pushpavanam}},\ }\bibfield  {title} {\enquote {\bibinfo {title}
  {Self-propulsion in 2d confinement: Phoretic and hydrodynamic
  interactions},}\ }\href@noop {} {\bibfield  {journal} {\bibinfo  {journal}
  {arXiv preprint arXiv:2104.00930}\ } (\bibinfo {year} {2021})}\BibitemShut
  {NoStop}%
\bibitem [{\citenamefont {Qin}\ \emph {et~al.}(2010)\citenamefont {Qin},
  \citenamefont {Xia},\ and\ \citenamefont {Whitesides}}]{qin2010_soft}%
  \BibitemOpen
  \bibfield  {author} {\bibinfo {author} {\bibfnamefont {D.}~\bibnamefont
  {Qin}}, \bibinfo {author} {\bibfnamefont {Y.}~\bibnamefont {Xia}}, \ and\
  \bibinfo {author} {\bibfnamefont {G.~M.}\ \bibnamefont {Whitesides}},\
  }\bibfield  {title} {\enquote {\bibinfo {title} {Soft lithography for micro-
  and nanoscale patterning},}\ }\href@noop {} {\bibfield  {journal} {\bibinfo
  {journal} {Nature Protocols}\ }\textbf {\bibinfo {volume} {5}},\ \bibinfo
  {pages} {491--502} (\bibinfo {year} {2010})}\BibitemShut {NoStop}%
\bibitem [{\citenamefont {Petit}\ \emph {et~al.}(2016)\citenamefont {Petit},
  \citenamefont {Polenz}, \citenamefont {Baret}, \citenamefont {Herminghaus},\
  and\ \citenamefont {B{\"a}umchen}}]{petit2016_vesicles-on-a-chip}%
  \BibitemOpen
  \bibfield  {author} {\bibinfo {author} {\bibfnamefont {J.}~\bibnamefont
  {Petit}}, \bibinfo {author} {\bibfnamefont {I.}~\bibnamefont {Polenz}},
  \bibinfo {author} {\bibfnamefont {J.-C.}\ \bibnamefont {Baret}}, \bibinfo
  {author} {\bibfnamefont {S.}~\bibnamefont {Herminghaus}}, \ and\ \bibinfo
  {author} {\bibfnamefont {O.}~\bibnamefont {B{\"a}umchen}},\ }\bibfield
  {title} {\enquote {\bibinfo {title} {Vesicles-on-a-chip: {{A}} universal
  microfluidic platform for the assembly of liposomes and polymersomes},}\
  }\href@noop {} {\bibfield  {journal} {\bibinfo  {journal} {European Physical
  Journal E}\ }\textbf {\bibinfo {volume} {39}},\ \bibinfo {pages} {59}
  (\bibinfo {year} {2016})}\BibitemShut {NoStop}%
\bibitem [{\citenamefont {Bradski}(2000)}]{bradski2000_opencv}%
  \BibitemOpen
  \bibfield  {author} {\bibinfo {author} {\bibfnamefont {G.}~\bibnamefont
  {Bradski}},\ }\bibfield  {title} {\enquote {\bibinfo {title} {The {{OpenCV
  Library}}},}\ }\href@noop {} {\bibfield  {journal} {\bibinfo  {journal} {Dr
  Dobbs J. Softw. Tools}\ }\textbf {\bibinfo {volume} {25}},\ \bibinfo {pages}
  {120--125} (\bibinfo {year} {2000})}\BibitemShut {NoStop}%
\bibitem [{\citenamefont {Crocker}\ and\ \citenamefont
  {Grier}(1996)}]{crocker1996_methods}%
  \BibitemOpen
  \bibfield  {author} {\bibinfo {author} {\bibfnamefont {J.~C.}\ \bibnamefont
  {Crocker}}\ and\ \bibinfo {author} {\bibfnamefont {D.~G.}\ \bibnamefont
  {Grier}},\ }\bibfield  {title} {\enquote {\bibinfo {title} {Methods of
  digital video microscopy for colloidal studies},}\ }\href@noop {} {\bibfield
  {journal} {\bibinfo  {journal} {Journal of Colloid and Interface Science}\
  }\textbf {\bibinfo {volume} {179}},\ \bibinfo {pages} {298--310} (\bibinfo
  {year} {1996})}\BibitemShut {NoStop}%
\bibitem [{\citenamefont {Thielicke}\ and\ \citenamefont
  {Stamhuis}(2014)}]{thielicke2014_pivlab}%
  \BibitemOpen
  \bibfield  {author} {\bibinfo {author} {\bibfnamefont {W.}~\bibnamefont
  {Thielicke}}\ and\ \bibinfo {author} {\bibfnamefont {E.}~\bibnamefont
  {Stamhuis}},\ }\bibfield  {title} {\enquote {\bibinfo {title} {{{PIVlab}}
  \textendash{} {{Towards User}}-friendly, {{Affordable}} and {{Accurate
  Digital Particle Image Velocimetry}} in {{MATLAB}}},}\ }\href@noop {}
  {\bibfield  {journal} {\bibinfo  {journal} {Journal of Open Research
  Software}\ }\textbf {\bibinfo {volume} {2}},\ \bibinfo {pages} {e30}
  (\bibinfo {year} {2014})}\BibitemShut {NoStop}%
\bibitem [{\citenamefont {Hokmabad}\ \emph {et~al.}(2020)\citenamefont
  {Hokmabad}, \citenamefont {Saha}, \citenamefont {{Agudo-Canalejo}},
  \citenamefont {Golestanian},\ and\ \citenamefont
  {Maass}}]{hokmabad2020quantitative}%
  \BibitemOpen
  \bibfield  {author} {\bibinfo {author} {\bibfnamefont {B.~V.}\ \bibnamefont
  {Hokmabad}}, \bibinfo {author} {\bibfnamefont {S.}~\bibnamefont {Saha}},
  \bibinfo {author} {\bibfnamefont {J.}~\bibnamefont {{Agudo-Canalejo}}},
  \bibinfo {author} {\bibfnamefont {R.}~\bibnamefont {Golestanian}}, \ and\
  \bibinfo {author} {\bibfnamefont {C.~C.}\ \bibnamefont {Maass}},\ }\bibfield
  {title} {\enquote {\bibinfo {title} {Quantitative characterization of
  chemorepulsive alignment-induced interactions in active emulsions},}\
  }\href@noop {} {\bibfield  {journal} {\bibinfo  {journal} {arXiv}\ ,\
  \bibinfo {pages} {2012.05170}} (\bibinfo {year} {2020})}\BibitemShut
  {NoStop}%
\end{thebibliography}
\end{document}